\documentclass[pra,twocolumn,showpacs,eqsecnum,superscriptaddress]{revtex4}

\usepackage{color}
\usepackage{multirow}
\usepackage{amsmath}
\usepackage{graphicx}
\usepackage{float}
\usepackage{verbatim}

\begin{document}


\newcommand{\bra}[1]{\langle{#1}|}
\newcommand{\ket}[1]{|{#1}\rangle}
\newcommand{\id}{\openone}
\newcommand{\eq}[1]{(\ref{#1})}
\newcommand{\vphi}{\varphi}
\renewcommand{\vec}[1]{\mathbf{#1}}
\newcommand{\bo}{\boldsymbol{(}}
\newcommand{\bc}{\boldsymbol{)}}

\newcommand{\sz}[1]{\sigma_{z_{#1}}}
\newcommand{\sx}[1]{\sigma_{x_{#1}}}
\newcommand{\sy}[1]{\sigma_{y_{#1}}}
\newcommand{\spp}[1]{\sigma_{+_{#1}}}
\newcommand{\smm}[1]{\sigma_{- _{#1}}}

\newcommand{\tsz}[1]{\tilde\sigma_{z_{#1}}}
\newcommand{\tsx}[1]{\tilde\sigma_{x_{#1}}}
\newcommand{\tsy}[1]{\tilde\sigma_{y_{#1}}}
\newcommand{\tspp}[1]{\tilde\sigma_{+_{#1}}}
\newcommand{\tsmm}[1]{\tilde\sigma_{-_{#1}}}

\newcommand{\sch}{Schr\"odinger}
\newcommand{\schs}{Schr\"odinger's}
\newcommand{\nn}{\nonumber}
\newcommand{\nl}{\nn \\ &&}
\newcommand{\dg}{^\dagger}
\newcommand{\rt}[1]{\sqrt{#1}\,}
\newcommand{\ito}{It\^o }
\newcommand{\str}{Stratonovich }
\newcommand{\erf}[1]{Eq.~(\ref{#1})}
\newcommand{\erfs}[2]{Eqs.~(\ref{#1}) and (\ref{#2})}
\newcommand{\erft}[2]{Eqs.~(\ref{#1}) -- (\ref{#2})}

\newcommand{\im}{\mathrm{Im}}

\newcommand{\wa}[1]{\omega_{a_{#1}}}
\newcommand{\twa}[1]{\tilde\omega_{a_{#1}}}
\newcommand{\wrr}{\omega_{r}}
\newcommand{\rf}{\mathrm{rf}}
\newcommand{\spec}{\mathrm{s}}

\newcommand{\w}[1]{\omega_{#1}}
\newcommand{\wdr}[1]{\omega_{d_{#1}}}

\newcommand{\ad}{a^\dag}
\newcommand{\aop}{a}
\newcommand{\veps}{\varepsilon}

\newcommand{\ccos}{\mathrm{c}}
\newcommand{\ssin}{\mathrm{s}}
\newcommand{\ej}[1]{E_\mathrm{J_{#1}}}
\newcommand{\ec}[1]{E_\mathrm{C_{#1}}}
\newcommand{\eel}[1]{E_\mathrm{el_{#1}}}
\newcommand{\vlc}{V_\mathrm{LC}}

\newcommand{\nc}{n_\mathrm{crit}}

\newcommand{\etal}{{\it et. al}}

\newcommand{\U}[1]{{\mathsf{#1}}}

\def\be{\begin{equation}}
\def\ee{\end{equation}}

\title{Quantum information processing with circuit quantum electrodynamics}
\date{\today}

\author{Alexandre Blais}
\affiliation{Departments of Applied Physics and Physics, Yale University, New Haven, CT 06520}
\affiliation{D\'epartement de Physique et Regroupement Qu\'eb\'ecois sur les Mat\'eriaux de Pointe, Universit\'e
de Sherbrooke, Sherbrooke, Qu\'ebec, Canada, J1K 2R1}
\author{Jay Gambetta}
\affiliation{Departments of Applied Physics and Physics, Yale University, New Haven, CT 06520}
\author{A.~Wallraff}
\affiliation{Departments of Applied Physics and Physics, Yale University, New Haven, CT 06520}
\affiliation{Department of Physics, ETH Zurich, CH-8093 Z{\"u}rich, Switzerland}
\author{D.~I.~Schuster}
\affiliation{Departments of Applied Physics and Physics, Yale University, New Haven, CT 06520}
\author{S.~M.~Girvin}
\affiliation{Departments of Applied Physics and Physics, Yale University, New Haven, CT 06520}
\author{M.~H.~Devoret}
\affiliation{Departments of Applied Physics and Physics, Yale University, New Haven, CT 06520}
\author{R.~J.~Schoelkopf}
\affiliation{Departments of Applied Physics and Physics, Yale University, New Haven, CT 06520}

\begin{abstract} 
We theoretically study single and two-qubit dynamics in the circuit QED architecture.  We focus on the current  experimental design [Wallraff {\it et al.}, Nature {\bf 431}, 162 (2004); Schuster {\it et al.}, Nature  {\bf 445}, 515 (2007)] in which  superconducting charge qubits are capacitively coupled to a single high-Q superconducting coplanar resonator.  In this system, logical gates are realized by driving the resonator with microwave fields.   Advantages of this architecture are that it allows for multi-qubit gates between non-nearest qubits and for the realization of gates in parallel, opening the possibility of fault-tolerant quantum computation with superconduting circuits.  In this paper, we focus on one and two-qubit gates that do not require moving away from the charge-degeneracy `sweet spot'. This is advantageous as it helps to increase the qubit dephasing time and does not require modification of the original circuit QED.  However these gates can, in some cases, be slower than those that do not use this constraint. Five types of two-qubit gates are discussed, these include gates based on virtual photons, real excitation of the resonator and a gate based on the geometric phase.   We also point out the importance of selection rules when working at the charge degeneracy point.
\end{abstract}


\maketitle

\section{Introduction}

Superconducting circuits based on Josephson junctions~\cite{devoret:2004,wendin:2006} are currently the most
experimentally advanced solid-state qubits.  The quantum behavior of these circuits has been experimentally
tested at the level of a single qubit~\cite{wallraff:2005,chiorescu:2003,vion:2002,martinis:2002,nakamura:99}
and of a pair of qubits~\cite{mcdermott:2005,majer:2005,Yamamoto:2003,Berkley:2003,pashkin:2003}.  The first
quantitative experimental study of entanglement in a pair of coupled superconducting qubits was recently
reported~\cite{steffen:2006b}.

In this paper, we theoretically study quantum information processing for superconducting charge qubits in
circuit QED~\cite{blais:2004,wallraff:2004,schuster:2005,wallraff:2005,schuster:2007,gambetta:2006}, focusing on
two-qubit gates.  In this system qubits are coupled to a high Q transmission line resonator which acts as a
quantum bus.  Coupling of superconducting qubits through a quantum bus has already been studied by
several authors and in different settings.  In particular, coupling using a lumped LC oscillator~\cite{makhlin:99,you:2002,you:2003b,zhou:2004,migliore:2005,liu:2005a,liu:2006,zagoskin:2006}, an extended 1D or 3D resonator~\cite{you:2003a,yang:2003,yang:2005,gywat:2006,paternostro:2006,wallquist:2006},  a current-biased Josephson junction acting as an anharmonic oscillator~\cite{blais:2003,plastina:2003,wang:2004,wei:2005,wallquist:2005} or using a mechanical oscillator~\cite{sornborger:2004,geller:2005,pritchett:2005} were studied. Here, we focus on circuit QED with
charge qubits~\cite{blais:2004} and consider the constraints of the {\em current} experimental 
design~\cite{wallraff:2004,schuster:2005,wallraff:2005,schuster:2007}.    As we will show, while this
architecture is simple, it allows for many different types of
qubit-qubit interactions.  These gates have the advantage that they can be realized between non-nearest qubits,
possibly spatially separated by several millimeters.  In addition to being interesting from a fundamental point
of  view, this is highly advantageous in reducing the complexity of multi-qubit
algorithms~\cite{blais:2001}.  Moreover, it also helps in reducing the error threshold required for reaching
fault-tolerant quantum computation~\cite{gottesman:2000}.  Furthermore,  some of the gates that will be
presented allow for parallel operations (i.e. multiple one and two-qubit gates acting simultaneously on different pairs
of qubits).  This feature is in fact a requirement for a fault-tolerance threshold to exist~\cite{aharonov:96},
and this puts circuit QED on the path for scalable quantum computation.

Another aspect addressed in this paper is the `quality' of realistic implementations of these gates.  To
quantify this quality, several measures, like the fidelity, have been proposed~\cite{poyatos:1997}.  A fair
evaluation and comparison of these measures for the different gates however requires extensive numerical
calculations including realistic sources of imperfections and optimization of the gate parameters.  In this
work, we will rather present estimates for the quality factor~\cite{vion:2002} of the gates as obtained from
analytical calculations.  Initial numerical calculations have showed that, in most situations, better results
than predicted by the analytical estimates can be obtained by optimization.  The quality factors presented here
should thus be viewed as lower bounds on what can be achieved in practice.

Five types of two-qubit gates will be presented.  First, we discuss in section~\ref{sec_variable_detuning} gates
that are based on tuning the transition frequency of the qubits in and out of resonance with the resonator by using
dc charge or flux bias.  As will be discussed, this approach is advantageous because it yields the fastest gates, 
whose rate given by the resonator-qubit coupling frequency.  A problem with this simple approach is that it
takes the qubits out of their charge-degeneracy `sweet spot', which can lead to a substantial  increase of their
dephasing rates~\cite{vion:2002}.  Moreover,  changing the qubit transition frequency over a wide range of
frequencies can be problematic if the frequency sweep crosses environmental resonances~\cite{simmonds:2004}.

To address these problems, we will focus in this paper on gates that do not require dc excursions away from the
sweet spot.  Requiring that there is no dc bias is a stringent constraint and the gates that are obtained will
typically be slower.  However,  the resulting gate quality factor can be larger because of the important gain in
dephasing time.  The first of these types of gates rely on virtual excitation of
the resonator (section~\ref{sec_virtual_gate}).  This type of approach was also discussed in
Refs.~\cite{blais:2004,gywat:2006} and here we will present various mechanisms to tune
this type of interaction.  The dispersive regime that is the basis of the schemes relying on virtual excitations
of the resonator can also be used to create probabilistic entanglement due to measurement.  This is discussed in
section~\ref{sec_entanglement_by_measurement}.   Next, we consider gates that are based on real photon
population of the resonator (section~\ref{sec_real_photons}).  For these gates, selection rules will set some
constraints on the transitions that can be used.  Finally, we discuss a gate based on the geometric phase
which was first introduced in the context of ion trap quantum
computation~\cite{garcia-ripoll:2003,garcia-ripoll:2005}. 

Before moving to two-qubit gates, we begin in section~\ref{sec_circuit_qed_review} with a brief review of
circuit QED and, in section~\ref{sec_review_1_qubit}, with a discussion of single-qubit gates.   A table
summarizing the expected rates and quality factors for the different gates is presented in the concluding
section.

\section{Circuit QED}
\label{sec_circuit_qed_review}

\subsection{Jaynes-Cummings interaction}
\label{sec_circuit_qed_review_JC}

\begin{figure}[tp]
\centering \includegraphics[width=1\columnwidth]{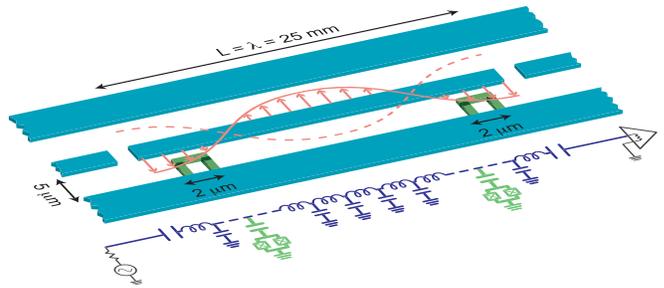} \caption{(Color online) Layout and lumped element
version of circuit QED.  Two superconducting charge qubit (green) are fabricated inside the superconducting 1D
transmission line resonator (blue).} \label{fig_circuit}
\end{figure}

In this section, we briefly review the circuit QED architecture first introduced in Ref.~\cite{blais:2004}
and experimentally studied in Refs.~\cite{wallraff:2004,schuster:2005,wallraff:2005,schuster:2007}. 
Measurement-induced dephasing was theoretically studied in
Ref.~\cite{gambetta:2006}.  As shown in Fig.~\ref{fig_circuit}, this system consists of a superconducting
charge qubit~\cite{bouchiat:1998,makhlin:2001,devoret:2004} strongly coupled to a transmission line
resonator~\cite{frunzio:2004}.      Near its resonance frequency $\omega_r$, the transmission line resonator can
be modeled as a simple harmonic oscillator composed of the parallel combination of an inductor $L$ and a
capacitor $C$.   Introducing the annihilation  (creation) operator $a^{(\dag)}$, the resonator can then be
described by the Hamiltonian
\be
H_{r} = \wrr \ad \aop,
\ee
with $\wrr = 1/\sqrt{LC}$ and where we have taken $\hbar =1$.  
Using this simple model, the voltage across the LC circuit (or, equivalently, on the
center conductor of the resonator) can be written as $\vlc = V^0_\mathrm{rms}(\ad+ \aop)$, where
$V^0_\mathrm{rms} = \sqrt{\hbar\wrr/2C}$ is the rms value of the voltage in the ground state.  An important
advantage of this architecture is the extremely small separation $b \sim 5\:\mu$m between the center conductor
of the resonator and its ground planes.  This leads to a large rms value of the electric field $E^0_\mathrm{rms}
= V^0_\mathrm{rms}/b \sim 0.2$~V/m for typical
realizations~\cite{wallraff:2004,schuster:2005,wallraff:2005,schuster:2007}.

Multiple superconducting charge qubits can be fabricated in the space between the center conductor and the
ground planes of the resonator.  As shown in Fig.~\ref{fig_circuit}, we will consider the case of two
qubits fabricated at the two ends of the resonator.  These qubits are sufficiently far apart that the direct
qubit-qubit capacitance is negligible.  Direct capacitive coupling of qubits fabricated inside a resonator was
discussed in Ref.~\cite{gywat:2006}. An advantage of placing the qubits at the ends of the resonator is the
finite capacitive coupling between each qubit and the input or output port of the resonator.  This can be used
to independently dc-bias the qubits at their charge degeneracy point.   The size of the direct capacitance must
be chosen in such a way as to limit energy relaxation and dephasing due to noise at the input/output ports. Some
of the noise is however still filtered by the high-Q resonator~\cite{blais:2004}.  We note, that recent design
advances have also raised the possibility of eliminating the need for dc bias altogether.~\cite{schuster:2007}

In the two-state approximation, the Hamiltonian of the $j$th qubit takes the form
\be
H_{q_j} = -\frac{\eel{j}}{2}\sx{j} - \frac{\ej{j}}{2}\sz{j}, \label{eq_H_qubit}
\ee
where $\eel{j} = 4\ec{j}(1-2n_{g_j})$ is the electrostatic energy and $\ej{j} = E^\mathrm{max}_\mathrm{J_j}
\cos(\pi \Phi_j/\Phi_0)$ is the Josephson coupling energy.  Here, $\ec{j} = e^2/2C_{\Sigma_j}$ is the charging
energy with $C_{\Sigma_j}$ the total box capacitance.  $n_{g_j} = C_{g_j}V_{g_j}/2e$ is the dimensionless gate
charge with $C_{g_j}$ the gate capacitance and $V_{g_j}$ the gate voltage.  $E^\mathrm{max}_\mathrm{J_j}$ is the
maximum Josephson energy and $\Phi_j$ the externally applied flux with $\Phi_0$ the flux quantum.  Throughout
this paper, the $j$ subscript will be used to distinguish the different qubits and their parameters.

With both qubits fabricated close to the ends of the resonator (antinodes of the voltage), the coupling to the
resonator is maximized for both qubits.  This coupling is capacitive and determined by the gate voltage $V_{g_j}
= V_{g_j}^\mathrm{dc} + \vlc$, which contains both the dc contribution $V_{g_j}^\mathrm{dc}$ (coming from a dc
bias applied to the input port of the resonator) and a quantum part $\vlc$.  Following
Ref.~\cite{blais:2004}, the Hamiltonian of the circuit of Fig.~\ref{fig_circuit} in the basis of the
eigenstates of $H_{q_j}$ takes the form
\begin{equation}
\begin{split}
H & =  \omega_r \ad a + \sum_{j=1,2} \frac{\wa{j}}{2}\sz{j}\\
& - \sum_{j=1,2} g_j \left( \mu_j - \ccos_j \sz{j} + \ssin_j\sx{j}  \right)\left(\ad+a\right),
\end{split}
\label{eq_H_2qubit_full}
\end{equation}
where $\wa{j} = \sqrt{\ej{j}^2+[4\ec{j}(1-2n_{g,j})]^2}$ is the transition frequency of qubit $j$ and $g_j =
e(C_{g,j}/C_{\Sigma,j})V^0_\mathrm{rms}/\hbar$ is the coupling strength of the resonator to qubit $j$.  For
simplicity of notation, we have also defined $\mu_j = 1-2n_{g,j}$, $\ccos_j = \cos\theta_j$ and
$\ssin_j=\sin\theta_j$, where $\theta_j = \arctan(\ej{j}/\ec{j}(1-2n_{g,j}))$ is the mixing
angle~\cite{blais:2004}.

When working at the charge degeneracy point $n_{g,j}^\mathrm{dc} = 1/2$, where dephasing is
minimized~\cite{vion:2002}, and neglecting fast oscillating terms using the rotating-wave approximation (RWA),
the above resonator plus qubit Hamiltonian takes the usual Jaynes-Cummings form~\footnote{Here we use $-g$
instead of $+g$.  We have dropped this unimportant minus sign in Ref.~\cite{blais:2004}.}
\be
H_\mathrm{JC} = \wrr \ad \aop + \sum_{j=1,2}\frac{\wa{j}}{2}\sz{j} - \sum_{j=1,2}  g_j
\left(\ad\smm{j}+\spp{j} \aop\right). \label{eq_Hjc}
\ee
This coherent coupling between a single qubit and the resonator was investigated experimentally in
Ref.~\cite{wallraff:2004,schuster:2005,wallraff:2005,schuster:2007}.  In particular, in
Ref.~\cite{wallraff:2005} high fidelity single qubit rotations were demonstrated.

\subsection{Damping}

Coupling to additional uncontrollable degrees of freedom leads to energy relaxation and dephasing in the system.
In the Born-Markov approximation, this can be characterized by a photon leakage rate $\kappa$ for the resonator,
an energy relaxation rate $\gamma_{1,j}$ and a pure dephasing rate  $\gamma_{\phi,j}$ for each qubit.  In the
presence of these  processes, the state of the qubit plus cavity system is described by a mixed state $\rho(t)$
whose evolution follows the master equation~\cite{walls-milburn}
\be
\begin{split}
\dot\rho = & -i[H,\rho] +\kappa\mathcal{D}[\aop]\rho
+\sum_{j=1,2}\gamma_{1,j}\mathcal{D}[\smm{j}]\rho
\\
&+\sum_{j=1,2}\frac{\gamma_{\varphi,j}}{2}\mathcal{D}[\sz{j}]\rho,
\end{split}
\label{eq_master_eq}
\ee
where $\mathcal{D}[\hat L]\rho = \left(2 L \rho L^\dag - L^\dag  L \rho - \rho   L^\dag  L\right)/2$ describes
the effect of the baths on the system.

\subsection{Typical system parameters}

In this section, we give realistic system parameters.  The resonator frequency $\wrr/2\pi$ will be assumed to be
between 5 and 10 GHz.  The qubit transition frequencies can be chosen anywhere between about 5 to 15 GHz, and
are tunable by applying a flux though the qubit loop.  In the schematic circuit of Fig.~\ref{fig_circuit}, both
qubits are affected by the externally applied field, but the effect on each qubit will depend on the qubit's
loop area.  Coupling strengths $g/2\pi$ between 5.8 and 100 MHz have been realized
experimentally~\cite{wallraff:2004,schuster:2007} and couplings up to 200 MHz should be feasible.

Rabi frequencies of 50 MHz where obtained with a sample of moderate coupling strength $g/2\pi =$ 17
MHz~\cite{wallraff:2005} and an improvement by at least a factor of two is realistic.

The cavity damping rate $\kappa$ is chosen at fabrication time by tuning the coupling capacitance between the
resonator center line and it's input and output ports.  Quality factors up to $Q\sim 10^6$ have been reported
for under-coupled resonators~\cite{day:2003,frunzio:2004}, corresponding to a low damping rate $\kappa/2\pi =
\wrr/2\pi Q \sim$ 5 KHz for a $\wrr/2\pi =$ 5 GHz resonator.  This results in a long photon lifetime $1/\kappa$
of 31 $\mu$s. To allow for fast measurement, the coupled quality factor can also be reduced by two or more
orders of magnitude.

Relaxation and dephasing of a qubit in one realization of this system were measured in Ref.~\cite{wallraff:2005}.  There, $T_1 =
7.3~\mu$s and $T_2$ = 500 ns were reported.  These translate to $\gamma_1/2\pi = 0.02$ MHz and $\gamma_\phi/2\pi
= (\gamma_2 - \gamma_1/2)/2\pi$ =  0.31 MHz.

\section{1-qubit gates}
\label{sec_review_1_qubit}

Single qubit gates are realized by pulses of microwaves on the input port of the resonator.  Depending on the
frequency, phase and amplitude of the drive, different logical operations can be realized.  External driving of
the resonator can be described by the Hamiltonian
\be
H_\mathrm{D} = \sum_{k}  \left(\veps_{k}(t)  \ad e^{-i \wdr{k}t}+ \veps_{k}^*(t)  \aop e^{+i
\wdr{k}t}\right), \label{eq_H_drive}
\ee
where $\veps_k(t)$ is the amplitude and $\omega_{d_k}$ the frequency of the $k$th external drive. Throughout
this paper, the $k$ subscript will be used to distinguish between the different drives and the drive-dependent
parameters.

For simplicity of notation, we first consider the situation where there is a single qubit and drive present.  We
will also assume that the qubit is biased at its optimal point and use the RWA.   The Hamiltonian describing
this situation is $H=H_\mathrm{JC}+H_\mathrm{D}$ with $j=k=1$.

Logical gates are realized with microwaves pulses that are substantially detuned from the resonator frequency.
With a high-Q resonator, this means that a large fraction of the photons will be reflected at the input port.  
To get useful gate rates, we thus work with large amplitude driving fields.  In this situation, quantum fluctuations in 
the drive are very small with respect to the drive amplitude and the drive can be considered, for all practical purposes, 
as a classical field.  In this case, it is convenient to displace the field operators using the time-dependent displacement 
operator~\cite{scully}
\be
D(\alpha) = \exp\left(\alpha \ad - \alpha^* \aop\right). \label{eq_Displacement_op}
\ee
Under this transformation, the field $a$ goes to $a+\alpha$ where $\alpha$ is a c-number representing the 
classical part of the field.

The displaced Hamiltonian reads
\be
\begin{split}
\tilde H
& = D^\dag(\alpha)HD(\alpha) - i D^\dag(\alpha)\dot D(\alpha) \\
&= \wrr \ad \aop +\frac{\wa{}}{2}\sz{} -  g \left(\ad\smm{}+\spp{} \aop\right)\\
&-  g\left(\alpha^* \smm{} +\alpha \spp{}  \right),
\end{split}
\ee
where we have chosen $\alpha(t)$ to satisfy
\be
\dot\alpha = -i \omega_r\alpha - i \varepsilon(t) e^{-i\wdr{} t}. \label{eq_alpha_displacement}
\ee
This choice of $\alpha$ is made so as to eliminate the direct drive on the resonator Eq.~\eq{eq_H_drive}  from the effective Hamiltonian.

In the case where the drive amplitude $\veps$ is independent of time, and by moving to a frame rotating at the
frequency $\omega_d$ for both the qubit and the field operators, we get
\be
\tilde H  = \Delta_r \ad \aop +\frac{\Delta_a}{2}\sz{} -  g \left(\ad\smm{}+\spp{} \aop\right) +
\frac{ \Omega_R}{2} \sx{}, \label{eq_Rabi}
\ee
where we have dropped any transient in $\alpha(t)$. In the above expression, we have defined $\Delta_r =
\omega_r-\wdr{}$ which is the detuning of the cavity from the drive, $\Delta_a = \omega_a-\wdr{}$ the detuning
of the qubit transition frequency from the drive and $\Omega_R $  is the Rabi frequency:
\be
\Omega_R = 2\frac{\veps g}{\Delta_r}.
\ee
In the limit where $\Delta_r$ is large compared with the resonator half-width $\kappa/2$, the average photon
number in the resonator can be written as $\bar n \approx (\veps/\Delta_r)^2$.  In this case, the Rabi frequency
takes the simple form $\Omega_R \approx 2g\sqrt{\bar n}$ expected from the Jaynes-Cummings model.

We note that the effect of damping can be taken into account by performing the
transformation~\eq{eq_Displacement_op} on the master equation~\eq{eq_master_eq} rather than on Schr\"odinger's
equation.  For completeness, this is done in appendix~\ref{sec_displace_master}.  Since in this paper we are
interested in the qubit dynamics under coherent control rather than measurement, we will be working in the
regime where $\Delta_r > \kappa$ and as such can safely ignore the effect of $\kappa$ on $\Omega_R$. For a
detailed discussion of measurement in this system, see Ref.~\cite{gambetta:2006}.

\subsection{On-resonance: bit-flip gate}
\label{sec_1_qubit_gate_x}


For quantum information processing, it is more advantageous to work in the dispersive regime where $\Delta = \omega_a - \omega_r$) is much bigger than the coupling $g$.  One advantage of this regime is that the pulses aimed at controlling the qubit are far detuned from the resonator frequency and are thus not limited in speed by its high quality factor.  Another advantage is that the high quality resonator filters noise at the far detuned qubit transition frequency and effectively enhances the qubit lifetime~\cite{blais:2004}.

To take into account that we are working in this dispersive regime,  we eliminate the direct qubit-resonator coupling by using the transformation
\be
U = \exp\left[\frac{g}{\Delta}(\ad\smm{}-\aop\spp{})\right]. \label{eq_dispersive_U_1q}
\ee
Using the Hausdorff expansion to second order in the small parameter $\lambda = g/\Delta$
\be
e^{-\lambda X} H e^{\lambda X} = H + \lambda [H,X]  + \frac{\lambda^2}{2!}[[H,X],X]+\cdots \label{eq_Hausdorff}
\ee
with $X=(\ad\smm{}-\aop\spp{})$, yields~\cite{blais:2004}
\be
\begin{split}
H_x
&\approx  \Delta_r \ad \aop +\frac{1}{2}\left(\Delta_a+2\chi\left[\ad\aop + \frac{1}{2}\right]\right)\sz{} +  \frac{ \Omega_R}{2} \sx{}\\
&\approx  \Delta_r \ad \aop +\frac{\tilde\Delta_a}{2}\sz{} +  \frac{\Omega_R}{2} \sx{},
\end{split}
\label{eq_H_x}
\ee
where we have defined $\chi = g^2/\Delta$ and $\tilde\Delta_a = \twa{} - \wdr{}$ with $\twa{} =\wa{} + \chi$.
Since the resonator is driven far from the frequency band $\omega_r\pm\chi$ where cavity population can be
large, we have that $\langle\ad\aop\rangle\sim 0$ (this is because we are working in a displaced frame with 
respect to the resonator field).  As a result, we have therefore dropped the ac-Stark shift in
the second line of the above expression.

By choosing $\tilde\Delta_a = 0$, the above Hamiltonian generates rotations around the $x$ axis at a rate
$\Omega_R$.  These Rabi oscillations have already been observed experimentally in circuit QED with close to unit
visibility~\cite{wallraff:2005}.  Changing $\tilde\Delta_a$, $\Omega_R$ and the phase of the
drive can be used to rotate the qubit around any axis on the Bloch sphere~\cite{collin:2004}.


In the situation where many qubits are fabricated in the resonator and have different transition frequencies,
the qubits can be individually addressed by tuning the frequency of the drive accordingly.  It should therefore
be possible to individually control several qubits in the circuit QED architecture.

\subsection{Off-resonance: phase gate}
\label{sec_1q_gates_off_res}

It is useful to consider the situation where the drive is sufficiently detuned from the qubit that it cannot
induce transitions, but is of large enough amplitude to significantly ac-Stark shift the qubit transition
frequency due to virtual transitions.  To obtain an effective Hamiltonian describing this situation, we start by
adiabatically eliminating the effect of direct transitions of the qubit due to the drive.  This is be done by
using on Eq.~\eq{eq_Rabi} the transformation
\be
U=\exp\left[\beta^*\spp{}-\beta\smm{}\right] \label{eq_U_off_qubit}
\ee
to second order in the small parameter $\beta = \Omega_R/2\Delta_a$.  In a second step, we again take into
account the fact that the qubit is only dispersively coupled to the resonator by using the transformation of
Eq.~\eq{eq_dispersive_U_1q} to second order.  These two sequential transformation yield
\be
H_z  \approx\Delta_r \ad \aop + \frac{1}{2}\left(\tilde\Delta_a +
\frac{1}{2}\frac{\Omega_R^2}{\Delta_a}\right)\sz{}. \label{eq_phase_gate_1q}
\ee
The last term in the parenthesis is an off-resonant ac-Stark shift caused by virtual transitions of the qubit.
This shift can be used to realize controlled rotation of the qubit about the $z$ axis.  The rate of this gate
can be written in terms of the average photon number $\bar n$ inside the resonator as  $\sim2g\sqrt{\bar
n}\times(\Omega_R/2\Delta_a)$.  To get fast rotations, one must therefore choose large values of the coupling
constant $g$ and large $\bar n$ while keeping the ratio $ \Omega_R/2\Delta_a$ small to prevent real transitions.

Finally, it is important to point  out that in the situation where multiple qubits are present inside the
resonator, each qubit will suffer a frequency shift when other qubits are driven.  These frequency shifts will
have to be taken into account or canceled by additional drives.

\subsection{Coherent control vs.\@ measurement}
\label{sec_control_vs_measurement}

\begin{figure}[h]
\centering \includegraphics[width=1\columnwidth]{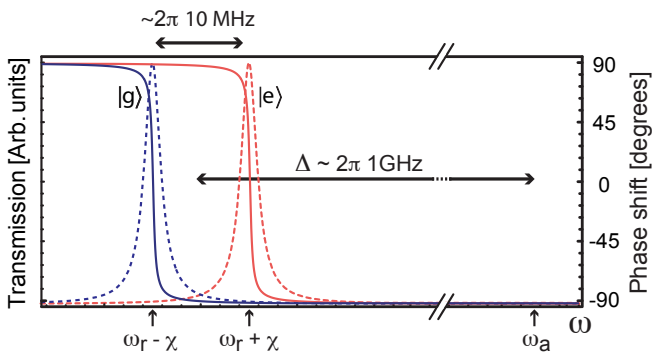} \caption{(Color online)
Resonator transmission (dashed lines) and corresponding phase shifts (full lines) for the two qubit states
(blue: ground, red: excited).  The numbers are calculated using $g/2\pi \sim$ 100 MHz and $g/\Delta \sim 0.1$.
At the far detuned frequencies required for single qubit gates, the phase shift of the resonator field does not
depend on the qubit state.  In this case, there is negligible entanglement between the resonator and the qubit.}
\label{fig_measurement_vs_gate}
\end{figure}

As mentioned above, in the dispersive regime, driving the cavity close to its resonance frequency leads to a
measurement of the qubits.  As discussed in Refs.~\cite{blais:2004,gambetta:2006}, this is due to entanglement
of the qubit with the resonator field generated by the term $\chi\ad\aop\sz{}$ of~Eq.~\eq{eq_H_x}.  Indeed,
because of this term, the resonator frequency is pulled to $\wrr\pm\chi$ depending on the state of the qubit.
The possible resonator transmissions, corresponding to the qubit in the ground (blue) or excited (red) state,
are shown (dashed lines) along with the corresponding phase shifts (full lines) in
Fig.~\ref{fig_measurement_vs_gate}.

As is seen from this figure, only around $\wrr\pm\chi$ is there significant phase shift and/or resonator
transmission change for the information rate about the qubit's state to be large at the resonator
output~\cite{gambetta:2006}.  In other words, only around these frequencies is entanglement between the
resonator and the qubit significant.  However, when coherently controlling the qubit using the flip and phase
gates discussed above, the resonator is irradiated far from $\wrr\pm\chi$.  As shown in
Fig.~\ref{fig_measurement_vs_gate}, since we are working in the dispersive regime where $|\Delta|\gg g$, there
is no significant phase difference in the resonator output between the two states of the qubit at these very
detuned frequencies.  As a result, there is no significant unwanted entanglement with the resonator when
coherently controlling the qubit.

An additional benefit of working at these largely detuned irradiation frequencies is that the resonator is only
virtually populated and the speed of the gates is not limited by the high Q of the resonator.  These two aspects
lead to high quality single qubit gates~\cite{wallraff:2005}.

\begin{figure}[tbp]
\centering \includegraphics[width=1\columnwidth]{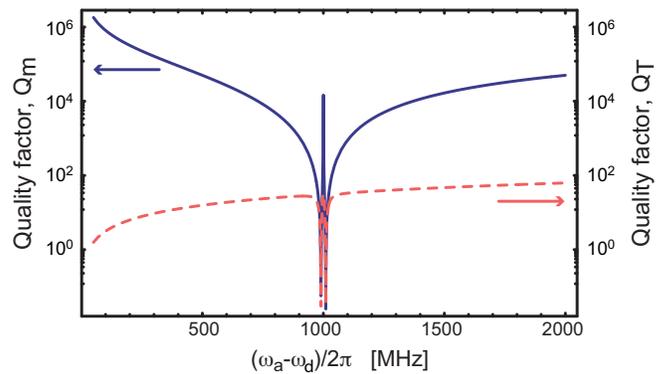} \caption{(Color online) Quality factor of the
phase gate, with respect to the measurement-induced dephasing rate $\Gamma_m$ (blue, full line) and with respect
to $\gamma_2+\Gamma_m$ (red, dashed line) where $1/\gamma_2$ = 500 ns.  This is plotted vs.\@ the detuning
between the control drive and the qubit transition frequency.  The amplitude $\veps$ of the drive is chosen such
that $\Omega_R/2\Delta_a = 0.1$ for all values of the detuning.  The other parameters are  $g/2\pi=100$ MHz,
$g/\Delta=0.1$ and $\kappa/2\pi = 0.5$ MHz.  The discontinuity in $\Gamma_m$ around
$(\wa{}-\omega)/2\pi=\Delta=1000$ MHz is due to the fact at that frequency, the drive is on resonance with the
resonator and therefore corresponds to a measurement.  This is not a region where a phase gate would be
operated.} \label{fig_phase_gate_rate}
\end{figure}

The above discussion can be made more quantitative by introducing the rate $\Gamma_m$ of dephasing induced by
the control drive (corresponding to measurement-induced dephasing)~\cite{gambetta:2006}:
\be
\Gamma_m =\frac{\kappa\chi^2(\bar n_+ +  \bar n_-)}{(\kappa/2)^2+\Delta_r^2+\chi^2} \label{eq_Gamma_m}
\ee
where
\be
\bar n_\pm = \frac{\veps^2}{(\kappa/2)^2+(\Delta_r\pm\chi)^2},
\ee
is the steady-state average photon number inside the resonator for a qubit  in the ground ($-$) or excited ($+$)
state.  In practice, this rate will always be much smaller than the intrinsic dephasing rate $1/T_2$ of the
qubit. For example, for the bit-flip gate, a  Rabi rate of $\Omega_R/2\pi=100$ MHz with $g/2\pi = $ 100 MHz and
$g/\Delta = 0.1$ yields a measurement-induced dephasing time $1/\Gamma_m$ of the order of a few milliseconds.
Clearly, this is not a limitation in practice.  This is illustrated for the phase gate in
Fig.~\ref{fig_phase_gate_rate} where the quality factor
\begin{equation}
Q_\phi = \frac{\Omega_R^2/2\Delta_a}{2\Gamma_\phi} \label{eq_quality_factor}
\end{equation}
is plotted as a function of the detuning of the drive with respect to the qubit transition frequency.  In this
figure, the full blue line is the quality factor ($Q_m$) calculated using  the measurement-induced dephasing
rate $\Gamma_m$ and the dashed red line the quality factor ($Q_T$) using the total rate $\Gamma_T =
\gamma_2+\Gamma_m$ assuming a dephasing time $1/\gamma_2$ of 500 ns.

For the phase gate, a dephasing time of $1/\gamma_2$ = 500 ns with  a rate of $\Omega_R^2/4\pi\Delta_a = 40$ MHz
at a detuning $(\wa{}-\wdr{})/2\pi=2000$ MHz yields a quality factor of $\sim 60$.  For the bit-flip gate, a
Rabi rate of 100 MHz yields a quality factor of $ \sim 157$.

\subsection{AC-dither: phase gate}
\label{sec_1q_gates_ac}

Another approach to produce a single-qubit phase gate, is to take advantage of the quadratic dependence of the
qubit transition frequency on the gate voltage (or flux) to shift the qubit transition frequency.  This can be
done by modulation of these control parameters at a frequency that is adiabatic with respect to the qubit
transition frequency.

Focussing on the single qubit Hamiltonian~\eq{eq_H_qubit}, we take $n_g(t) = n_g^\mathrm{dc} + n_d(t)$, where
$n_d(t) = n_g^\text{ac}\sin(\omega_\text{ac}t)$ is a modulation of the gate voltage that is slow compared to the
qubit transition frequency.  In this situation, it is useful to move to the adiabatic basis.  The relation
between the original ($\sigma_j$) and adiabatic ($\tilde\sigma_j$)  basis Pauli operators is given by
\be
\begin{split}
\sz{} &= \cos\Theta(t)\tsz{} - \sin\Theta(t)\tsx{},\\
\sx{} &= \sin\Theta(t)\tsz{} + \cos\Theta(t)\tsx{}, \label{eq_adiabatic_trans}
\end{split}
\ee
where $\Theta(t) = \arctan[E_\mathrm{el}(t)/E_\mathrm{j}]$.  In this basis, the qubit Hamiltonian reads
\be
\bar H =-\frac{\omega_a^\mathrm{ad}(t)}{2}\tsz{},
\ee
with $\omega_a^\mathrm{ad}(t) = \sqrt{E_\mathrm{J}^2 + \{4E_\mathrm{C}[1-2n_g^\mathrm{dc} - 2 n_d(t)]\}^2}$ the
instantaneous splitting.  Because of the quadratic dependence with gate charge, the average part of the qubit
splitting is larger than its bare value.  For example, setting $n_g = 1/2$ and assuming small dither amplitude,
we obtain
\be
\omega_a^\mathrm{ad} \approx \ej{} + 16 \frac{\ec{}^2}{\ej{}}(n_g^\text{ac})^2 \label{eq_dither_shifted_wa}
\ee
which should be compared to the bare value $\ej{}$.  Here, we have dropped  terms rotating at
$2\omega_\text{ac}$ and higher order in the dither amplitude.  Voltage ac-dither can therefore be used to
blue-shift the qubit transition frequency (flux dither around the flux sweet would cause a red shift).  As will
be discussed below, this can also be used to couple qubits when the dither frequency is larger than the coupling
strength (but still slow with respect to the qubit transition frequency).

\begin{figure}[t]
\centering \includegraphics[width=1\columnwidth]{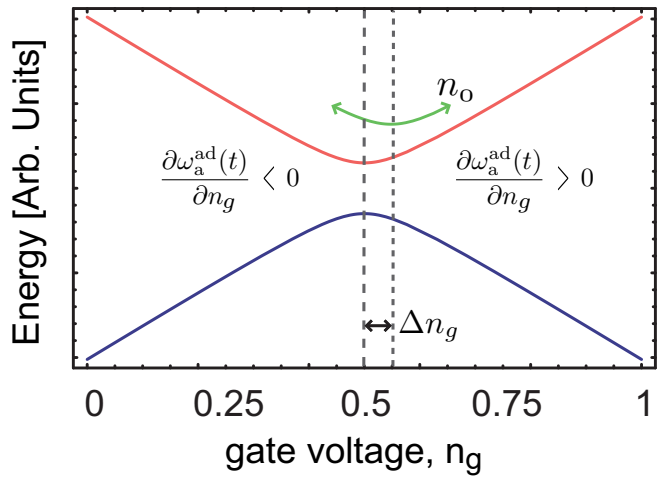} \caption{(Color online) Refocussing due to
ac-dither.  The qubit energy splitting is shown as a function of gate charge.  Ac-dither acts as refocussing by
sampling the positive and negative dependence of $\omega_a^\mathrm{ad}(t)$ with $n_g$.} \label{fig_ac_dither_refocussing}
\end{figure}

Because ac-dither acts as a continuous spin-echo~\cite{slichter}, it can be realized with minimal dephasing of
the qubit.  It therefore appears to be more advantageous than dc bias of the control parameters.  For example,
if the qubit is dc-biased $\Delta n_g$ away from the charge degeneracy point, $n_g^\mathrm{dc} = 1/2+\Delta
n_g$, noise $\delta n_g(t)$ in the bias will cause dephasing due to fluctuations $(\partial \wa{}/\partial
n_g)\delta n_g(t)$ in the qubit transition frequency.  As illustrated in Fig.~\ref{fig_ac_dither_refocussing},
if the dc-offset $\Delta n_g$ is small compared to the dither amplitude $n_g^\text{ac}$, then under dither, both
signs of $\partial \wa{}/\partial n_g$ will be probed leading to (partial) cancelation of the unwanted
fluctuating phase.

This cancelation can be seen more explicitly by assuming small excursions away from the charge degeneracy point
such that $\Theta(t)\approx -\lambda(\Delta n_g+n_g^\text{ac}\sin\omega_\text{ac} t)$, where we have defined
$\lambda \equiv 8E_\mathrm{C}/E_\mathrm{J}$.  Assuming that the qubit is not too deep in the charge regime such that $\lambda n_g^\text{ac}$ is small, we
expand to first order in $\Delta n_g$ and $n_g^\text{ac}$ to obtain
\be
\begin{split}
\bar H  & \approx
-\frac{\omega_a^\mathrm{ad}(t)}{2}\tsz{}\\
&+ 4 E_\mathrm{C}\delta n_g(t)
\left[-\lambda\Delta n_g J_0\left(\lambda n_g^\text{ac}\right) - 2 J_1 (\lambda
n_g^\text{ac})\sin\omega_\text{ac} t \right]\tsz{}
\\
&
+  4 E_\mathrm{C}\delta n_g(t) \left[J_0\left(\lambda n_g^\text{ac}\right) - 2\lambda\Delta n_g J_1(\lambda
n_g^\text{ac})\sin\omega_\text{ac} t \right]\tsx{},
\end{split}
\label{eq_dither_H}
\ee
where the $J_n(z)$ are Bessel functions of the first kind.  In this expression, we have dropped higher order Bessel functions and have added the gate charge
noise $\delta n_g(t)$ to $n_g(t)$.

The second term in Eq.~\eq{eq_dither_H}, proportional to $\tsz{}$, leads to pure dephasing $T_\phi$ while the
last term leads to mixing of the qubit. Focusing on pure dephasing, we obtain from the Golden rule
\be
\begin{split}
\frac{1}{T_\phi} & \approx \frac{1}{2} \left(\frac{64 E_\mathrm{C}^2}{E_\mathrm{J}} \right)^2 \Big\{ \Delta
n_g^2 S_{\delta n_g}(0)
\\
&
\quad + \frac{(n_g^\text{ac})^2}{4}\left[S_{\delta n_g}(-\omega_\text{ac}) + S_{\delta n_g}(+\omega_\text{ac})
\right] \Big\}, \label{eq_T_phi_dither}
\end{split}
\ee
where $S_{\delta n_g}(\omega_\text{ac})$ is the spectral density of the charge noise.

For both the dc and ac bias, the dephasing rate $1/T_\phi$ increases linearly with the amplitude of the shift of
the transition.  However, blue shifting of the qubit using ac-dither $n_g^\text{ac}$ produces less dephasing
than using the static bias $\Delta n_g$ provided that the dither frequency is much higher than the
characteristic frequency of the noise so that
\be
S_{\delta n_g}(\pm\omega_\text{ac}) \ll S_{\delta n_g}(0).
\ee
This approach should therefore efficiently protect the qubit from low frequency (i.e. $1/f$) noise.  Assuming
$\Delta n_g = 0$ and using Eq.~\eq{eq_dither_shifted_wa}, the quality factor of this gate can be estimated as
$Q\sim(\ej{}/64\ec{}^2)^2/(S_{\delta n_g}(-\omega_\text{ac}) + S_{\delta n_g}(+\omega_\text{ac}))$.  This type
of stabilization of logical gate by ac-fields was also studied in Ref.~\cite{fonseca-romero:2005}.

\section{Direct coupling by variable detuning}
\label{sec_variable_detuning}

In the layout of Fig.~\ref{fig_circuit}, qubit-qubit interaction must be mediated by the resonator.  It is
therefore reasonable to expect that the limiting rate on two-qubit gates be the coupling strength $g$.  Gates at
this rate can be implemented by taking the qubits in and out of resonance with the resonator frequency.  When a
qubit is far off resonance, it only dispersively couples to the resonator through the coupling
$\chi\ad\aop\sz{}$, where $\chi = g^2/\Delta$.  This interaction can be made small by working at large detunings
$\Delta$.  In this situation there is no significant qubit-resonator interaction.  The interaction is turned on
by tuning the qubit transition frequency back in resonance with the resonator.  In this case, vacuum Rabi
flopping at the frequency $2g$ will entangle the qubit and the resonator.  It is know from ion-trap quantum
computing~\cite{cirac:95,childs:2000}, and is further discussed in Section~\ref{sec_gates}, how to use this type
of interaction to mediate qubit-qubit entanglement.

Tuning of the qubit transition frequency could be realized by applying flux pulses through the individual qubit
loop.  This would require adding flux lines in proximity to the qubits.  Voltage bias using individual bias
lines could also be used, but this likely introduce more noise than flux bias.  Moreover, in both cases, this
will take the qubits away from their sweet spot, possibly increasing their dephasing rates~\cite{vion:2002}.
Alternatively, Wallquist {\it et al.} have suggested that a similar tuning of $\Delta$ can be realized by
fabricating a resonator whose frequency is itself tunable~\cite{wallquist:2006}.   While this is a promising
idea, one drawback is that any noise in the parameter controlling the resonator frequency will lead to dephasing
of photon superpositions, lowering the expected gate quality.

In addition to dc-bias, it is possible to use any of the rf approaches discussed in the previous section to tune
the qubit transition frequency.  Moreover, the FLICFORQ protocol~\cite{rigetti:2005}, discussed in the next
section and in appendix~\ref{sec_appendix_flicforq_1q} could also be used.  As shown in
appendix~\ref{sec_appendix_flicforq_1q}, this would yield qubit-resonator coupling at the rate $g/2$. However,
for these approaches to be useful here, very large rf amplitudes would be required to cover the large range of
frequency needed to turn on and off the qubit-resonator interaction.  This is especially true in the presence of many qubits fabricated in the same resonator.  A FLICFORQ-type protocol~\cite{rigetti:2005} was also suggested for flux qubits
coupled to a LC oscillator in Ref.~\cite{liu:2006}.  In this case, the authors considered quantum computation in the basis of the qubit dressed by a rf-drive directly applied directly to the flux qubit.

In summary, this type of gate relying on tuning of the qubit or resonator frequency is advantageous because it
operates at the optimal rate $g$.  However, it requires either additional bias lines and extra dephasing or
large amplitude rf-pulses.  In the next sections, we will focus on gates that do not require additional tuning
but only rely on rf-drive of the resonator of more moderate amplitudes.  While these gates will be typically
slower than the gates discussed here, they can be implemented without modification of the original circuit QED
design and do not suffer from the above problems.  Gates relying on direct tuning of the qubit transition
frequency will be further discussed elsewhere.

\section{2-qubit gates: virtual qubit-qubit interaction}
\label{sec_virtual_gate}

In this section, we expand the discussions of Ref.~\cite{blais:2004} on two-qubit gates using virtual
excitations of the resonator.  To minimize dephasing, we will work with both qubits at charge degeneracy
($\Delta n_g=0$).   In the rotating wave approximation, the starting point is therefore~Eq.~\eq{eq_Hjc}.   To
avoid excitation of the resonator, we assume that both qubits are strongly detuned from the resonator
$|\Delta_j|=|\wa{j}-\wrr|\gg g_j$.  In this situation, we adiabatically eliminate the resonant Jaynes-Cummings
interaction using the transformation
\be
U = \exp\left[\frac{g_1}{\Delta_1}(\ad\smm{1}-a\spp{1})+\frac{g_2}{\Delta_2}(\ad\smm{2}-a\spp{2})\right].
\label{eq_U_dispersive_2q}
\ee
To second order in the small parameters $g_j/\Delta_j$, this
yields~\cite{blais:2004,sorensen:99,imamoglu:1999,zheng:2000}
\be
\begin{split}
H_{\rm 2q} \approx \: & \omega_r \ad a
+ \sum_{j=1,2} \frac{\twa{j}}{2}\sz{j}\\
& + \frac{g_1g_2 (\Delta_{1}+\Delta_{2})}{2\Delta_{1}\Delta_{2}} \left(\spp{1}\smm{2}+\smm{1}\spp{2}\right),
\end{split}
\label{eq_simple_dispersive_H_2q}
\ee
where, as in section~\S\ref{sec_1_qubit_gate_x}, we have assumed that the cavity is in the vacuum state and have
taken $\twa{j} = \wa{j} + \chi_j$.  It is simple to generalize the above expression for an arbitrary number of
qubits coupled to the same mode of the resonator.  The last term in the above Hamiltonian describes swap of the
qubit states through virtual interaction with the resonator.  Evolution under this Hamiltonian for a time
$t=\pi\Delta_1\Delta_2/2g_1g_2(\Delta_1+\Delta_2)$ will generate a $\sqrt{i\mathrm{SWAP}}$
gate~\cite{blais:2004}.  This gate, along with the  single qubit gates discussed in section
\ref{sec_review_1_qubit}, form a universal set for quantum computation~\cite{barenco:95}.

In the situation where the qubits are strongly detuned from each other, energy conservation suppresses this
flip-flop interaction.  This is most easily seen by going to a frame rotating at $\twa{j}$ for each qubit.  In
this frame, when the qubits are strongly detuned, the interaction term is oscillating rapidly and averages out.
In this situation, the effective qubit-qubit interaction is for all practical purposes turned off.  On the other
hand, for $\twa{1}=\twa{2}$, the interaction term does not average and the interaction is effective.

To turn on and off this virtual interaction, it is necessary to change the detuning between the qubits.  There
are several ways to do this in the circuit QED architecture.  One possible approach is to directly change the
transition frequency of the qubits using, as described in section~\ref{sec_circuit_qed_review}, using flux or
voltage as control parameters.  However, as can be seen from~eq.~\eq{eq_T_phi_dither}, moving the gate charge
away from the sweet spot will rapidly increase the dephasing rate~\cite{vion:2002,ithier:2005}.

\subsection{Off-resonant ac-Stark shift}

The off-resonant ac-Stark shift discussed in section~\ref{sec_1q_gates_off_res} provides another way to tune the
qubits in and out of resonance.  In this situation, one must generalize the Hamiltonian
of~Eq.~\eq{eq_simple_dispersive_H_2q} to include off-resonant microwave fields.  This is done in
appendix~\ref{sec_appendix_eff_H} in the presence of three independent fields and two qubits.  Two of the
fields, of amplitudes $\veps_1$ and $\veps_2$, are used to coherently control the state of the qubits while the
third, of amplitude $\veps_3$ is used to readout the state of the qubits.

In this section, it is sufficient to take into account a single drive $\veps_1$, assumed to be strongly detuned
from any resonances.   The resulting effective Hamiltonian [5th term of Eq.~\eq{eq_H_eff_all}] contains the swap
term already obtained in the absence of coherent drive in Eq.~\eq{eq_simple_dispersive_H_2q}.  The effect of the
drive is to shift the qubit transition to
\be
\wa{j}'' =  \wa{j} +  \frac{\Omega_{R_j}^2}{2\Delta_{a_j}} +
2\frac{g_j^2}{\Delta'_j}\left(\langle\ad\aop\rangle+\frac{1}{2}\right), \label{eq_acStark_swap}
\ee
where $\Delta'_j =  \wa{j} +  \Omega_{R_j}^2/2\Delta_{a_j} - \wrr$ is the shifted detuning also entering in  the
renormalized swap rate.

\begin{figure}[t]
\centering \includegraphics[width=1\columnwidth]{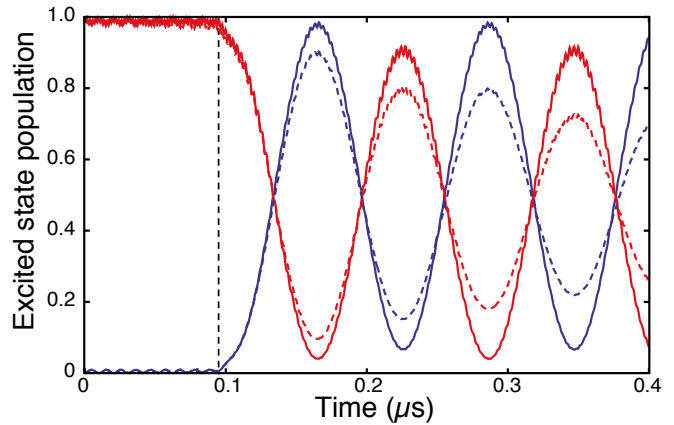} \caption{(Color online) Master equation
simulation of the excited state population of qubit one (blue) and qubit two (red) as a function of time.  Full
lines:  no relaxation and dephasing.  Dashed lines: finite relaxation and dephasing. The qubits are initially
detuned form each other with no drive applied.  A drive, bringing the shifted qubit transition frequencies in
resonance, is applied at the time indicated by the vertical dashed line.  At this time, the two-qubit state
start to swap.  The parameters are:  $g_1/2\pi=$ 80 MHz, $g_2/2\pi=$ 120 MHz, $\wrr/2\pi=$ 5 GHz, $\wa{1}/2\pi=$
7.1 GHz, $\wa{2}/2\pi=$ 7.0 GHz, $\wdr{1}/2\pi=$ 5200.14 MHz.  For the dashed lines, cavity decay was taken as
$\kappa/2\pi=0.22$ MHz and the qubits decay rates were taken to be identical and equal to $\gamma_1/2\pi=0.02$
MHz and $\gamma_\phi/2\pi=0.31$ MHz.} \label{fig_ac_Stark_flipflop}
\end{figure}

The strategy is then to chose $\wa{1}\neq \wa{2}$ such that the swap gate is effectively turned off in the idle
state.  The interaction is turned on by choosing a drive amplitude and frequency such that $\wa{1}''=\wa{2}''$.
This condition can be satisfied with a single drive provided that $g_1\neq g_2$. A master equation simulation of
this off-resonant ac-Stark tuning is illustrated in Fig.~\ref{fig_ac_Stark_flipflop}.

\subsection{AC-Dither}

In the same way as the off-resonant ac-Stark shift, ac-dither discussed in section~\ref{sec_1q_gates_ac} can be
used to effectively tune on resonance a pair of qubits (ac-dither being a low frequency version of the
off-resonant ac-Stark shift).  In this situation, the dither frequency must be faster than the swap rate between
the qubit but still adiabatic with respect to the qubit transition frequencies.  Moreover, similarly to the
off-resonant ac-Stark shift, both qubits will be blue shifted by the ac-dither.  The qubits can nevertheless be
tuned in resonance by taking advantage of their different direct capacitance to the input or output ports (as
discussed in section~\ref{sec_circuit_qed_review_JC}) and by using different $E_\mathrm{C}/E_\mathrm{J}$ ratios.

\subsection{FLICFORQ}
\label{sec_flicforq}

Another approach to tune off-resonant qubits is to use the so-called FLICFORQ protocol (Fixed LInear Couplings
between Fixed Off-Resonant Qubits)~\cite{rigetti:2005}.  In this protocol, one is interested in tuning the
effective interaction between a pair of qubits that are interacting through a fixed linear coupling.  The
coupling is assumed to be off-diagonal in the computational basis such that when the qubits are detuned, the
coupling is only a small perturbation and can be safely neglected.  The interaction is effectively turned on by
irradiating each qubit at its respective transition frequency and choosing the amplitude of the drives such that
one of the Rabi sidebands for one qubit is resonant with a Rabi sideband of the other qubit.  In this situation,
the effective coupling becomes first order in the bare coupling.

In circuit QED, FLICFORQ can be used in the dispersive regime to couple a qubit to the resonator or to couple a
pair of qubits together.  As shown in appendix~\ref{sec_appendix_flicforq_1q}, the resonance condition for the
first case is $\Delta = -\Omega_R$ and this leads to the effective qubit-resonator coupling
$(g/2)(\ad\smm{}+\aop\spp{})$, at the charge degeneracy point and in the RWA.  The interaction is first order in
the coupling $g$, but of reduced strength.

Similarly, two qubits that are dispersively coupled to the resonator and detuned from each other can be coupled
using FLICFORQ.  For simplicity, we again work at the charge degeneracy point for both qubits and use the
rotating wave approximation on the qubit-resonator couplings.  To turn on the interaction, two coherent drives of
frequency $\wdr{j}\approx\wa{j}$ are used.   Since the qubits are irradiated at their transition frequency, the
results of appendix~\ref{sec_appendix_eff_H} cannot be used directly.  The corresponding effective Hamiltonian
is derived in appendix~\ref{sec_appendix_flicforq_2q}.  At one of the possible sideband matching conditions and
in a quadruply rotating frame (see Appendix~\ref{sec_appendix_flicforq_2q}), the resulting effective Hamiltonian
is
\be
H_\mathrm{FF} \approx \wrr \ad \aop + \frac{g_1g_2(\Delta'_1+\Delta'_2)}{16\Delta'_1\Delta'_2}
\left(\sy{1}\sy{2}+\sz{1}\sz{2}\right),
\ee
where $\Delta'_j = \omega_{a_j}' -\omega_r$ with
\be
\wa{1(2)}'  = \wa{1(2)}  + 2 \frac{\Omega_{R_{12(21)}}^2}{\Delta_{a_{12(21)}}}.
\ee
the shifted qubit frequency.  In this expression, we have introduced $\Omega_{R_{jk}} = 2
g_j\veps_k/(\omega_r-\wdr{k})$ the Rabi frequency of qubit $j$ with respect to drive $k$ and
$\Delta_{a_{jk}}=\wa{j}-\wdr{k}$.  This effective qubit-qubit coupling is sufficient, along with single qubit
gates, for universal quantum computation.  Moreover, as expected from Ref.~\cite{rigetti:2005},  in the
FLICFORQ protocol,  the qubit-qubit coupling strength is reduced by a factor of 8 with respect to the bare
coupling strength.

\subsection{Fast entanglement at small detunings}
\label{sec_small_detuning}

The rates for the two-qubit gates described above are proportional to $g\times(g/\Delta)$, where $g/\Delta$ must
be small for the dispersive approximation to be valid.  An advantage of the dispersive regime is that the
resonator is only virtually populated and therefore photon loss is not a limiting factor.  The drawback is that
unless $g$ is large, dispersive gates can be slow.

It is interesting to see whether this rate can be increased by working at smaller detunings.  In this situation,
residual entanglement with the resonator can lead to reduced fidelities. As an example, we take for simplicity
$g_1=g_2=g$ and $\wa{1}=\wa{2}=\wa{}$. Choosing
\be
\Delta = \frac{2\sqrt{2}g}{\sqrt{4m^2-1}},
\ee
where $m\ge1$ is an integer, one can easily verify that starting from $\ket{ge0}$ or $\ket{eg0}$ at $t=0$, the
qubits are in a maximally entangled state and the resonator in the state $\ket{0}$ after a time
\be
T = \frac{\pi}{\Delta}.
\ee
Two-qubit entanglement can therefore be realized in a time $\sim 1/g$ and with a small detuning $\Delta \sim g$,
without suffering from spurious resonator entanglement when starting from $\ket{ge0}$ or $\ket{eg0}$.

It is also simple to verify that $\ket{gg0}$ only picks up a phase factor after time $T$.  However, starting
from $\ket{ee0}$ leads to leakage to $\ket{gg2}$ at time $T$ and therefore to unwanted entanglement with the
resonator.  As a result, while the simple procedure described here is not an universal 2-qubit primitive for
quantum computation, it can nevertheless lead to qubit-qubit entanglement at a rate which is close to the cavity
coupling rate $g$.  We will describe in Section~\ref{sec_gates} an approach based on
Ref.~\cite{childs:2000} to prevent this type of spurious resonator-qubit entanglement.

Since the cavity is populated by real rather than virtual photons during this procedure, it is important to
estimate photon loss. Starting from $\ket{ge0}$ or $\ket{eg0}$, the maximum photon number in the cavity is given
by $n_\text{max} =(4m^2-1)/8m^2$.  The worst case scenario for the rate of photon loss is then $\Gamma =
n_\text{max}\kappa$. The gate quality factor, considering only photon loss, is therefore $Q\sim 2\pi
T^{-1}/\Gamma = 32\sqrt{2} gm^2/[\kappa(4m^2-1)^{3/2}]$.   For $g/2\pi = 200$~MHz, $\kappa/2\pi=1$~MHz and
$m=1$, this yields a large quality factor of $Q\sim 1700$.

\subsection{Summary}
\label{sec_virtual_gates_summary}

The gates discussed in this section (apart from Sec.~\ref{sec_small_detuning}) rely on virtual population of the
resonator.  A disadvantage of these gates is that they will typically be slow.  All of them roughly go as
$g\times(g/\Delta)$ where $g/\Delta$  is a small parameter.  Taking $g/\Delta\sim0.1$ and $g/2\pi = 200$ MHz, we
see that the rates of these gates will realistically not exceed 20 MHz.  Although not very large, this rate
nevertheless exceeds substantially the typical decay rates $\gamma_1$, $\gamma_\phi$ and $\kappa$ of circuit
QED~\cite{wallraff:2004,schuster:2005,wallraff:2005} and should be sufficiently large for the experimental
realization of these ideas.  An advantage of these virtual gates is however that, since the resonator field is
only virtually populated, the gates do not suffer from photon loss.  As a result, these gates could still be
realized with a resonator of moderate $Q$ factor (which is advantageous for fast
measurement~\cite{wallraff:2005}).

It is also interesting to point out that, in the situation where there are more than two qubits fabricated in
the resonator, the same approach can be used to entangle simultaneously two or more pairs of qubits.   This is
done, for example, by taking $\wa{1} = \wa{2} \neq \wa{3} = \wa{4}$ while still in the dispersive regime.  It is
simple to verify that this corresponds to two entangling gates acting in parallel on the two pairs of qubits.
This type of classical parallelism is an important requirement for a fault-tolerant threshold to
exist~\cite{aharonov:96}.

Finally, we point out that the dispersive coupling can also be used to couple $n>2$ qubits simultaneously.  This
is done by tuning the $n$ qubits in resonance with each other but all still in the dispersive regime.  This
leads to multi-qubit entanglement in a single step.

\section{Conditional entanglement by measurement}
\label{sec_entanglement_by_measurement}

As discussed in section~\ref{sec_control_vs_measurement} and in more detail in
Ref.~\cite{blais:2004,wallraff:2004,schuster:2005,wallraff:2005,gambetta:2006}, measurement can be
realized in this system by taking advantage of the qubit-state dependent resonator frequency pull.  In the
presence of a single qubit, the resonator pull is $\chi\sz{}$ and becomes $\sum_j^n\chi_j\sz{j}$ in the n-qubit
case.  If the pulls $\chi_j$ are different for all $j$'s and large enough with respect to the resonator
linewidth $\kappa$, each of the different n-qubit states
$\{\ket{ggg...g},\ket{egg...g},\ket{gegg...g},\ldots,\ket{eee...e}\}$
pulls the cavity frequency by a different amount. In this situation,  it should be possible to realize single
shot QND measurements of the n-qubit state. In a test of Bell inequalities, this multi-qubit readout capability
would offer a powerful advantage over separate measurement of each qubit.  Indeed, in the latter case the
effective readout fidelity would be the product of the individual readout fidelities.

The situation is also interesting in the case where the pulls $\chi$ are equal.  For example, in the two qubit
case, when $\chi_1=\chi_2=\chi$ (and taking $\wa{1}=\wa{2}=\wa{}$ for simplicity), the dispersive Hamiltonian of
equation~\eq{eq_simple_dispersive_H_2q} can be written as
\be
\begin{split}
H_{\rm 2q} & \approx \left[\omega_r +\chi (\sz{1}+\sz{2})\right] \ad a + \sum_{j=1,2}
\frac{1}{2}\left(\wa{}+\chi\right)\sz{j}
\\
& + \chi \left(\spp{1}\smm{2}+\smm{1}\spp{2}\right).
\end{split}
\label{eq_2q_dispersive_chi}
\ee
While this Hamiltonian is not QND for measurement of $\sz{j}$, it is QND for measurement of $(\sz{1}+\sz{2})$
since $[(\sz{1}+\sz{2}),H_{\rm 2q}]=0$.

More interestingly, in this situation, the states $\{\ket{ge},\ket{eg}\}$ while they may have different 
Lamb shifts have the same cavity pull. This implies that they cannot be distinguished by this measurement. 
The consequence of this observation can be made more explicit by rewriting~\eq{eq_2q_dispersive_chi} as
\be
\begin{split}
H_{\rm 2q} &= \wrr \ad\aop
+\chi \left( \ket{\psi_+}\bra{\psi_+} - \ket{\psi_-}\bra{\psi_-}\right) \\
& + \frac{1}{2}\left[\wa{}+2 \chi \left(\ad\aop+\frac{1}{2}\right)\right] \left(\ket{ee}\bra{ee} -
\ket{gg}\bra{gg}\right),
\end{split}
\ee
where $\ket{\psi_\pm}=\left(\ket{ge}\pm\ket{eg}\right)/\sqrt{2}$ are Bell states.  As a result, the projection
operators corresponding to measurement of the cavity pull are $\Pi_g=\ket{gg}\bra{gg}$, $\Pi_e=\ket{ee}\bra{ee}$
and $\Pi_\mathrm{Bell}=\ket{\psi_+}\bra{\psi_+}+\ket{\psi_-}\bra{\psi_-}$, with
$\sum_{k=g,e,\mathrm{Bell}}\Pi_k=1$.

An initial state $\ket{\psi_i}$ will thus, with probability  $\bra{\psi_i}\Pi_\mathrm{Bell}\ket{\psi_i}$,
collapse to a state of the form $\ket{\psi_f}=c_+\ket{\psi_+}+c_-\ket{\psi_-}$. Since the Bell states
$\ket{\psi_\pm}$ are eigenstates of $H_{\rm 2q}$, further evolution will keep the projected state in this
subspace.  For certain unentangled initial states [e.g. $(\ket{g}+\ket{e})\otimes(\ket{g}+\ket{e})/2$, 
which is created by $\pi/2$ pulses on each qubit], the
state after measurement will be a maximally entangled state.  As a result, conditioned on the measurement of zero
cavity pull, Bell states can be prepared.  This type of entanglement by measurement for solid-state
qubits was also discussed by Ruskov and Korotkov~\cite{ruskov:2003}.  We also point out that 
entanglement by measurement and feedback was studied in Ref~\cite{sarovar:2005a}.

\section{2-qubit gates: qubit-qubit interaction mediated by resonator excitations}
\label{sec_real_photons}

In this section, we consider gates that actively use the resonator as a means to transfer information between
the qubits and to entangle them.  More precisely, we will take advantage of the so-called red and blue sideband
transitions.  We first start by a very brief overview of ion-trap quantum computing to show the similarities and
differences with circuit QED and then present various protocols adapted to circuit QED.  We discuss the
realization of quantum gates based on these sideband transitions in section~\ref{sec_gates}.

\subsection{Ion-trap quantum computation}

Sideband transitions are used very successfully in ion-trap quantum
computation~\cite{haffner:2005,leibfried:2005}.  This approach to ion-trap quantum computing was first discussed
by Cirac and Zoller~\cite{cirac:95} and, more recently, was adapted to two-level atoms by Childs and
Chuang~\cite{childs:2000}.  Following closely Ref.~\cite{childs:2000} the Hamiltonian describing a trapped ion
is
\be
H = \wrr \ad \aop + \frac{\wa{}}{2} \sz{} - \lambda  \cos[\eta (\ad+\aop)-\omega t]\sx{}, \label{eq_H_ion}
\ee
where $\omega_r$ is the frequency of the relevant mode of oscillation of the ion in the trap, $\omega$ is the
laser frequency, $\eta = \sqrt{k^2/2Nm\omega_r}$ is the Lamb-Dicke parameter and $\lambda$ is the amplitude of
the magnetic field produced by the driving laser~\cite{childs:2000}.  Assuming $\eta\ll 1$ and choosing the
laser frequency $\omega = \wa{}+n\wrr$, the only part of the interaction term that is not rapidly oscillating
and thus does not average out is
\be
H_\text{n-blue} \approx \frac{\lambda\eta^n}{2n!} \left(a^{\dag n} \spp{} + a^n \smm{} \right).
\ee
For $\omega = \omega_a - n \omega_r$, we get
\be
H_\text{n-red} \approx \frac{ \lambda\eta^n}{2n!} \left(a^{\dag n} \smm{} + a^n \spp{} \right).
\ee
These correspond to blue and red n-phonon sideband transitions, respectively.  It is known that these
transitions, along with single qubit gates, are universal for quantum computation on a chain of
ions~\cite{childs:2000}.

We first note that the rate of the sideband transitions scales with $\lambda$ which itself depends on the
(variable) laser amplitude.  In circuit QED, we will see that the (fixed) qubit-resonator coupling $g$ takes the
role of the parameter $\lambda$.  Moreover, while the Hamiltonian \eq{eq_H_ion} allows for all orders of
sideband transitions, we will see that when working at the charge degeneracy point, the symmetry of the circuit
QED Hamiltonian only allows sideband transitions of even orders.  Finally, it is interesting to point out that
the Hamiltonian~\eq{eq_H_ion} also describes a superconducting qubit magnetically coupled to a resonator.  In
that situation, the qubit would be fabricated at an anti-node of current in the resonator.  The zero-point
fluctuations  of the current generate a field that couples to the qubit loop.  For example, a superconducting
charge qubit magnetically coupled to the transmission line resonator would have in its Hamiltonian a term of the
form $\lambda\cos\left[\eta(\ad+\aop)+\bar\theta\right] \sx{}$, where $\lambda = \ej{}$ and $\eta = \pi M
I^0_\mathrm{rms} /\Phi_0$ with $M$ the mutual inductance between the center line of the transmission line,
$I^0_\mathrm{rms}$ the rms value of the vacuum fluctuation current on the center line in the ground
state~\cite{you:2003a,yang:2006b}, $\Phi_0$ is the superconducting flux quantum and $\bar\theta$ is controlled
by the dc flux bias.

An important difference between trapped ions and the superconducting flux qubit analog is that in the first case
$\lambda$ scales with the laser amplitude while, in the second case, it is equal to the Josephson energy.  The
latter can only be so large in practice and is difficult to tune rapidly.
Finally, we point out that solid-state qubits, in contrast to ions, have a natural spread in transition frequencies.  
This allows to address the qubits individually with global pulses, without requiring
 individual bias lines  to tune them.

\subsection{Sideband transitions in Circuit QED}
\label{sec_cqed_sidebands}

Our starting point to study sideband transitions in circuit QED is the two-qubit Hamiltonian of
Eq.~\eq{eq_H_2qubit_full}.  Here, we keep the gate charge dependence and do not initially make the rotating wave
approximation.  The effective Hamiltonian describing the sideband transitions in circuit QED is obtained in
appendix~\ref{sec_appendix_eff_H}.  As discussed above, in this appendix we consider the presence of two qubits
and three coherent drives.  Two of these drives, of frequency $\wdr{1,2}$ and amplitude $\veps_{1,2}$, are used
to drive the sideband transitions.  The third drive, of frequency $\wdr{3}$ and amplitude $\veps_3$, is used to
measure the state of the system.  For simplicity of notation, in this section we focus on a single qubit coupled
to the resonator and drop the $j$ index.

\begin{figure}[t]
\centering \includegraphics[width=1\columnwidth]{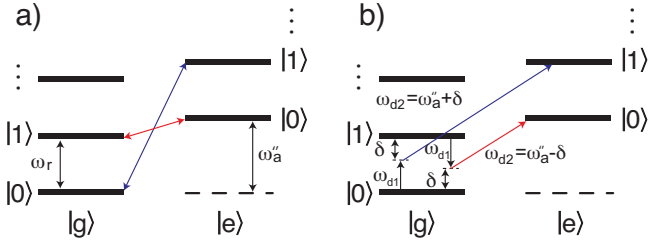} \caption{(Color online)  Red and blue sideband
transitions. a) One-photon transitions. b) Two-photon transitions.} \label{fig_red_blue}
\end{figure}

The red and blue sideband transitions, illustrated in Fig.~\ref{fig_red_blue}, are given by the last two lines
of Eq.~\eq{eq_H_eff_all}.  These correspond to single and two-photon sideband transitions, respectively.  Higher
order transitions are neglected due to their small amplitudes.  We rewrite these terms here in a more explicit
form.  This is done by going to a frame rotating at $\wrr$ for the resonator and at the shifted frequency
$\wa{j}''$ for the qubit, where
\be
\wa{}'' =  \wa{} +  \frac{\Omega_{R_{1}}^2}{2\Delta_{a_{1}}} +  \frac{\Omega_{R_{2}}^2}{2\Delta_{a_{2}}} +
2\frac{g^2}{\Delta'}\left(\langle\ad\aop\rangle+\frac{1}{2}\right). \label{eq_wapp}
\ee
Here we keep the contribution of $\langle\ad\aop\rangle$ to the frequency shift in order to take into account
the presence of the measurement drive $\veps_3$.  When evaluating $\langle\ad\aop\rangle$, it is important to
remember that, $\aop^{(\dag)}$ is in the displaced frame defined in appendix~\ref{sec_appendix_eff_H}.
Setting $\wdr{k^{(')}} = \wa{}'' + \wrr$ the only term that does not average to zero due to rapid oscillations
is the third line of~Eq.~\eq{eq_H_eff_all} which gives
\be
H_\mathrm{r1} = - \ccos g\frac{\Omega_{R_{k}}}{\Delta_{a_{k}}} \left( \spp{}\aop + \smm{}\ad  \right),
\label{eq_Heff_1_photon_red}
\ee
where $k=1$ or 2.  Following the notation introduced in Sec.~\ref{sec_circuit_qed_review_JC}, we use $\ccos =
\cos\theta$ and $\ssin = \sin\theta$ with $\theta$ the mixing angle.  The above Hamiltonian corresponds to the
one-photon red sideband transition.  Alternatively, taking $\wdr{k^{(')}} = \wa{}'' - \wrr$,  we obtain the
one-photon blue sideband transition
\be
H_\mathrm{b1} = - \ccos g \frac{\Omega_{R_{k}}}{\Delta_{a_{k}}} \left( \spp{}\ad + \smm{}\aop  \right).
\label{eq_Heff_1_photon_blue}
\ee

On the other hand, the last line of~\eq{eq_H_eff_all} will not average to zero if the drive frequencies are
chosen such that
\be
\wdr{k} \pm \wdr{k^{'}} = \wa{}'' + \wrr \label{eq_red_blue_freq_2photon}
\ee
or
\be
2 \wdr{k^{(')}} = \wa{}'' + \wrr. \label{eq_red_blue_freq_2photon_bis}
\ee
With these choices of drive frequencies, we obtain
\be
\begin{split}
H_\mathrm{r2} & =  \ssin g \frac{\Omega_{R_{1}}}{2\Delta_{a_{1}}} \frac{\Omega_{R_{2}}}{2\Delta_{a_{2}}}
\left(\spp{}\aop + \smm{}\ad\right),\\
H_\mathrm{b2} & =  \ssin g \frac{\Omega_{R_{1}}}{2\Delta_{a_{1}}} \frac{\Omega_{R_{2}}}{2\Delta_{a_{2}}}
\left(\spp{}\ad + \smm{}\aop\right).
\end{split}
\label{eq_Heff_2_photon}
\ee
corresponding to two-photon red and blue sideband transitions, respectively.  These two-photon transitions are
illustrated in Fig.~\ref{fig_red_blue}b).  

Because of the dependence of Eqs.~\eq{eq_Heff_1_photon_red} and \eq{eq_Heff_1_photon_blue} on the cosine of the
mixing angle $\theta$, the first order red and blue sideband transitions are forbidden at the charge degeneracy
point.  As discussed in appendix~\ref{sec_symmetry}, this can be linked to the symmetry of the Jaynes-Cummings
Hamiltonian.    Since it is more advantageous to work at the sweet spot to minimize dephasing, these first order
transitions therefore appear to be of limited interest for coherent control in circuit QED.   A similar selection rule,
for flux qubits coupled to a LC oscillator, was noted in Ref.~\cite{liu:2005a}.  There, it was suggested to work
with single-photon sidebands by biasing the flux qubit close to degeneracy, but not exactly at the sweet spot.
Moreover, selection rules for a flux qubit irradiated with classical microwave signal were also studied in 
Ref.~\cite{liu:2005}. We also point out that, since the
frequencies corresponding to these first order transitions ($\wa{j}''\pm\wrr$) are in practice very detuned from
the resonator frequency $\wrr$, signals at these frequencies would be mostly reflected at the input port of the
resonator.  Very large input powers would therefore be require to compensate the attenuation.

\begin{figure}[tbp]
\centering \includegraphics[width=1\columnwidth]{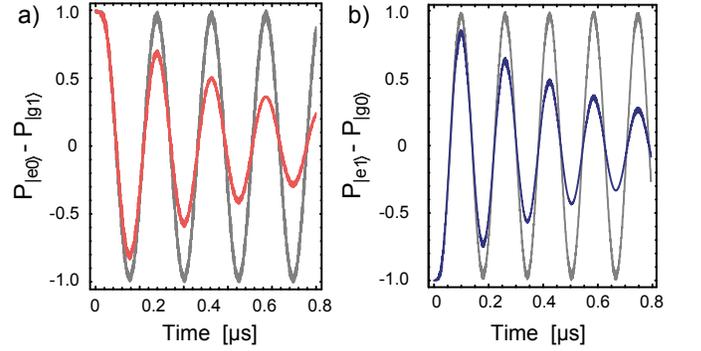} \caption{ (Color online) (a) Red and (b)
blue coherent sideband oscillations.  The system parameters are $\wa{}/2\pi $ = 7 GHz, $\wrr/2\pi = 5$ GHz,
$g/2\pi$= 100 MHz.  For the color curves, the damping is $\kappa/2\pi = 0.22$ MHz, $\gamma_1/2\pi$ = 0.02 MHz
and $\gamma_\phi/2\pi = $ 0.31 MHz.  These last two values are taken from the measured value reported in
Ref.~\cite{wallraff:2005}.  For the gray curves, damping was set to zero.  The value of the drive frequencies
and amplitudes are: a) $\wdr{1}/2\pi$ = 4593.59 MHz, $\wdr{2}/2\pi$ = 6650.605 MHz, $\veps_1/2\pi$ = 841.91 MHz
and $\veps_2/2\pi$ = 843.417 MHz,  b) $\wdr{1}/2\pi$ = 4598.39 MHz, $\wdr{2}/2\pi$ = 7476.47 MHz, $\veps_1/2\pi$
= 1241.51 MHz and $\veps_2/2\pi$ = 1252.56 MHz.} \label{fig_sidebands_oscillations}
\end{figure}

On the other hand, because of their $\sin\theta$ dependance, the two-photon transitions have maximal amplitude
at the sweet spot.  Moreover, since in this case we require the sum or the difference of the drive frequencies
to match the sideband conditions, these frequencies can be chosen such that there is  minimal attenuation (while
still avoiding measurement-induced dephasing~\cite{blais:2004,schuster:2005,gambetta:2006}).  However, the rate
for these 2-photon transitions is of order $g\times(\Omega_R/2\Delta_a)^2$, which is $g$ times the square of a
small parameter.  As is discussed in section~\ref{sec_virtual_gates_summary}, obtaining large rates will require
large coupling strength $g$.

To realize logical gates based on the red and blue sidebands, these transitions must be coherently driven.  The
corresponding simulated coherent oscillations are illustrated in Figure~\ref{fig_sidebands_oscillations} for the
two-photon red (a) and blue (b) sideband transitions.   In these numerical calculations, the drive frequency
$\wdr{1}$ is chosen $\sim 400$ MHz away from the resonator frequency to avoid measurement-induced dephasing of
the qubit.  In a first step, the second drive frequency $\wdr{2}$ is then chosen using the condition of
Eq.~\eq{eq_red_blue_freq_2photon}.  Using simulated annealing~\cite{aarts}, the drive frequencies and the
corresponding drive amplitudes $\veps_{1(2)}$ are then varied to optimize the fidelity of the sideband
transitions.  The best (but not necessarily optimal) values obtained are given in the caption of
Fig.~\ref{fig_sidebands_oscillations}.  Using these parameters, we obtain a population transfer of 0.83 for the
red sideband and of 0.86 for the blue sideband (without damping, we obtain near perfect population transfer of
1.0 in both cases).  This relatively low population transfer is essentially due to the small dephasing time
$1/\gamma_2 = 500$ ns used here with respect to the slow rate $\sim 5$ MHz of the red and blue sideband
transitions.   It is interesting to point out that this value of the rate is about 4 times bigger than expected
from the perturbative estimates of Eq.~\eq{eq_Heff_2_photon}.  These estimates should be taken as lower bounds
which can be improved by numerical optimisation.

\subsection{One-photon sidebands at the sweet spot using ac-dither}
\label{eq_ac-dither_sideband}

In the previous section, we saw that the symmetry of the circuit QED Hamiltonian does not allow for first order
red and blue sideband transitions at the charge degeneracy point.  However, with ac-dither discussed in
section~\ref{sec_1q_gates_ac}, it is possible to take advantage of the small gate charge excursions away from
the sweet spot to obtain a one-photon transition while staying, on average, at the sweet spot.  This can be seen
as a low frequency version of the two-photon transitions.

To analyse this situation, we focus on a single qubit coupled to the resonator and in the presence of a single
coherent drive of amplitude $\veps$ and frequency $\wdr{}$.  At the charge degeneracy point
($n_g^\mathrm{dc}=1/2$) and with ac-dither on the voltage port, we get for the Hamiltonian:
\be
\begin{split}
H & =  \omega_r\ad a +\frac{\ej{}}{2}\sz{}+ \frac{A}{2}\cos(\omega_\text{ac} t)\sx{}\\
& + g\left(2n_g^\text{ac}\cos(\omega_\text{ac}t) + \sx{}\right)\left(\ad + a\right)\\
& + \veps(\ad e^{-i\wdr{} t} + \aop e^{+i\wdr{} t}), \label{eq_H_dither_bare}
\end{split}
\ee
where we have defined $A = 8 \ec{} n_g^\text{ac}$ and have taken the ac-gate bias to be
$n_g^\text{ac}\cos(\omega_\text{ac}t)$.  Assuming that the dither frequency is small with respect to the qubit
transition frequency but large compared to the coupling strength, $g\ll \omega_\text{ac} \ll \wa{}$, we move to
the adiabatic basis to obtain
\be
\begin{split}
&H = \omega_r\ad a + \frac{\wa{}^\mathrm{ad}(t)}{2}\tsz{}
+ \veps(\ad e^{-i\wdr{} t} + \aop e^{+i\wdr{} t})\\
&+ g\left[2n_g^\text{ac}\cos(\omega_\text{ac}t) + \sin\Theta(t) \tsz{} + \cos\Theta(t) \tsx{}\right] \left(\ad +
a\right),
\end{split}
\label{eq_H_dither_adiabatic}
\ee
where $\wa{}^\mathrm{ad}(t) = \left[\ej{}^2+A^2\cos^2(\omega_\text{ac}t)\right]^{1/2}$ is the instantaneous
qubit transition frequency.

Comparing with Eq.~\eq{eq_H_2qubit_full}, we have the following mapping
\be
\mu \mapsto -2n_g^\text{ac}\cos(\omega_\text{ac}t), \quad \ccos  \mapsto \sin\Theta(t), \quad \ssin  \mapsto
-\cos\Theta(t) \label{eq_mapping}
\ee
and it is possible to use directly all of the results obtained in the previous section.   The important point is
that we now get a $\tsz{}$ component even at the charge degeneracy point, which means that the red and blue
sidebands will be allowed to first order.  This is due do the small excursions away from degeneracy provided by
the ac-dither.

Assuming that the dither amplitude is small with respect to the bare qubit transition frequency $\ej{}$, we take
$\Theta(t) \approx A\cos(\omega_\text{ac}t)/\ej{}$ which will yield simple Bessel function FM modulation
sidebands for the qubit transition.
Following section \ref{sec_1q_gates_ac} and using the results of the previous section, we get the blue sideband
transition for $\wdr{} = \omega''_a+\omega_r \pm \omega_\text{ac}$
\be
H_\mathrm{b1_{ac}} = g \left(\frac{\Omega_R}{2\Delta_a}\right) J_1\!\left(\frac{8\ec{}
n_g^{ac}}{\ej{}}\right)\left(\ad\tspp{}+a\tsmm{}\right),
\ee
and for $\wdr{} = \omega''_a-\omega_r \pm \omega_\text{ac}$ the red blue sideband transition
\be
H_\mathrm{r1_{ac}} = g \left(\frac{\Omega_R}{2\Delta_a} \right) J_1\!\left(\frac{8\ec{}
n_g^{ac}}{\ej{}}\right)\left(a\tspp{}+\ad\tsmm{}\right).
\ee
Here we have focused on the first dither sideband $J_1(z)$ only.  With respect to the two-photon transitions of
Eq.~\eq{eq_Heff_2_photon}, we have simply replaced a factor of $\Omega_R/2\Delta_a$ by the ac-sideband
modulation $J_1(8\ec{} n_g^{ac}/\ej{})$.  Both of these quantities are smaller than unity, so whether this is
advantageous depends on the parameters of the system.  As an example, taking $\ec{} \sim 5$~GHz, $\ej{} \sim
6$~GHz~\cite{schuster:2005}, and a 10\% of $2e$ excursion for the dither, $n_g^{ac} = 0.1$, we have  $J_1(8\ec{}
n_g^{ac}/\ej{}) \sim 3/10$.

\subsection{CNOT from sideband transitions}
\label{sec_gates}

In this section, we show how to obtain non-trivial two-qubit gates from red and blue sideband transitions.  This
is included for completeness, with most of the results already known from Cirac and Zoller~\cite{cirac:95} for
three level atoms and from Childs and Chuang~\cite{childs:2000} for two level atoms.   Here, we follow closely
the results  and notation of Ref.~\cite{childs:2000}.

Following Ref.~\cite{childs:2000}, we introduce the unitary operators~\footnote{Note the change of
convention for the operators $\spp{}$ and $\smm{}$ from Ref.~\cite{childs:2000}.  Here, we explicitely work in
the $\{\ket{e},\ket{g}\}$ basis and $\spp{}$ takes $\ket{g}$ to $\ket{e}$ as usual.}
\begin{align}
\U{R}^+_j(\theta,\phi) &= \exp\left[ -i \frac{\theta}{2} (e^{-i\phi}\ad \spp{j} + e^{+i\phi} a \smm{j})\right],\\
\U{R}^-_j(\theta,\phi) &= \exp\left[ -i \frac{\theta}{2} (e^{-i\phi}\ad \smm{j} + e^{+i\phi} a \spp{j})\right].
\end{align}
$\U{R}^+_j(\theta,\phi)$ corresponds to a pulse on the blue-sideband for qubit $j$ and $\U{R}^-_j(\theta,\phi)$
the red-sideband.  Here, $\phi$ is the phase of the driving field.

In addition to the above resonator-qubit operations, we introduce the single-qubit unitary operators
corresponding to the effective Hamiltonians discussed in section~\ref{sec_review_1_qubit}.  We denote the single
qubit flip ($x$) and phase ($z$)  operators acting on qubit $j$ as:
\be
\U{R}^x_j(\theta) = \exp\left[- i \frac{\theta}{2} \sx{j}\right], \qquad \U{R}^z_j(\theta) = \exp\left[- i
\frac{\theta}{2} \sz{j}\right],
\ee
Another useful single qubit unitary is the Hadamard transformation (in the basis $\{\ket{e},\ket{g}\}$)
\be
\U{H}_j = \frac{1}{\sqrt{2}}
\begin{pmatrix}
1&1\\
1&-1
\end{pmatrix},
\ee
which can be obtained by a rotation at an angle between the $x$ and $z$ axis or, equivalently, by a sequence of
one-qubit gates
\be
\begin{split}
\U{H}_j & = \exp\left[-i \frac{\pi}{2}\left(\frac{\sx{j}+\sz{j}}{\sqrt{2}}\right)\right]\\
& = \U{R}^x_j(\pi/2)\U{R}^z_j(\pi/2)\U{R}^x_j(\pi/2).
\end{split}
\ee

Before building a universal two-qubit gate from these elementary operations, we first discuss a simpler protocol
to create conditional entanglement.  This protocol is based on Ref.~\cite{blais:2003} and relies on
entangling one qubit to the resonator and then transferring the entanglement to qubit-qubit correlations.  This
is realized by the sequence
\be
\U{R}^-_1(\pi,\phi)\U{R}^-_2(\pi/2,\phi)\U{R}^-_1(\pi,\phi), \label{eq_simple_sequence}
\ee
where the phase $\phi$ is arbitrary.  It is simple to verify that this sequence will leave $\ket{gg0}$ unchanged
but will create maximally entangled states when starting from $\ket{ge0}$ or $\ket{eg0}$.  However, the state
$\ket{ee0}$ will irreversibly leak out of the $\{\ket{0},\ket{1}\}$ photon subspace and the qubits will get
entangled with the resonator at the end of the pulse sequence.  Hence, while~\eq{eq_simple_sequence} is not an
universal two-qubit primitive, it can nevertheless be used to generate conditional entanglement.  This sequence
can also be realized with all blue sideband pulses.  In this case, $\ket{ee0}$ is left unchanged while starting
with $\ket{gg0}$ will create spurious entanglement with the resonator.

In the above sequence, the spurious entanglement occurs in the second step where the initial state $\ket{ee0}$
picks up a contribution from the two-photon state $\ket{gg2}$.  For this state, the evolution is ``faster'' by a
factor of $\sqrt{2}$ from evolution in the one-photon subspace.  Because of this, the last step in the sequence
cannot completely undo the qubit-resonator entanglement and leaves them partially entangled.  To solve this
problem, Childs and Chuang~\cite{childs:2000} introduced the qubit-resonator gate
\be
\U{P_j}
 =  \U{R}^+_j(-\pi/2,0) \U{R}^+_j(\pi\sqrt{2},-\pi/2) \U{R}^+_j(-\pi/2,0).
\ee
In the basis $\{\ket{e0},\ket{e1},\ket{g0},\ket{g1}\}$, $\U{P_j}$ takes the form $\U{P_j} =\mathrm{diag}
(1,e^{-i\pi/\sqrt{2}},e^{i\pi/\sqrt{2}},-1)$.  This gate entangles the qubit with the resonator (because of the
minus sign in the last element) but it does not lead to leakage into higher photon states.  Using this gate,
Childs and Chuang~\cite{childs:2000} proposed a sequence of red, blue and single-qubit operations that generates
a CNOT gate.

Here, building on~Eq.~\eq{eq_simple_sequence}, we note a simpler entangling two-qubit gate (in the basis
$\{\ket{ee},\ket{eg},\ket{ge},\ket{gg}\}$)
\be
\begin{split}
\U{U_\varphi} &= \: \U{R}^+_1(\pi,\phi) \U{P_2}
\U{R}^+_1(\pi,\phi)\\
 &=\: \mathrm{diag} (1,e^{i\pi/\sqrt{2}},-e^{-i\pi/\sqrt{2}},1).
\end{split}
\ee
Using this gate, it is possible to obtain a CNOT gate which relies only on single qubit unitaries and
blue-sideband transitions:
\be
\U{CNOT} = \U{R}^z_1([1+\sqrt{2}]\pi) \U{H}_2 \U{U_\varphi} \U{R}^z_2(-\pi/\sqrt{2}) \U{H}_2.
\label{eq_blue_CNOT}
\ee
Because it relies on $\U{P_j}$, this gate does not lead to unwanted qubit-resonator entanglement.

\subsection{Summary}

The gates presented in this section rely on real excitations of the resonator to mediate entanglement between
the qubits.  These gates are based on perturbation theory and are therefore relatively slow.  For example, for
the red and blue sideband oscillations studied in section~\ref{sec_cqed_sidebands}, we have found after
numerical optimization rates of $\sim$ 5 MHz.  With a dephasing time of 500 ns, this corresponds to a quality
factor of about 9, larger than what was expected from perturbation theory.  While this quality factor is not
large enough for large scale quantum computation, it is certainly enough to demonstrate the concept
experimentally.

Finally, a disadvantage of the gates based on real excitation of the resonator is that they are susceptible to
photon loss and therefore require relatively large Q resonators.  This conflicts with the requirement of fast
readout.

\section{Geometric phase gate}
\label{sec_push_gate}

The gates discussed in the previous sections were based on real or virtual transitions.  In the present section,
we discuss a different approach, based on the geometric phase.  This was already discussed in the context of
ion-trap quantum computing~\cite{garcia-ripoll:2003,garcia-ripoll:2005}.  This gate is based on the first term
of the second line of the Hamiltonian~\eq{eq_H_eff_all}:
\be
\tilde H = -\sum_{j=1,2}g_jB_j\sz{j}(\ad+\aop), \label{eq_push_H_simple}
\ee
where $B_j$ is given by Eq.~\eq{eq_B}.   Here, we work at the charge degeneracy point where $\ccos_{1,2}=0$.
Although the Hamiltonian~\eq{eq_push_H_simple} does not couple the qubits directly, it couples both qubits to
the resonator field $\aop$.  By using a time dependent drive on the resonator, and hence displacing the field
$\aop$ in a controlled manner, it is possible to induce indirect qubit-qubit coupling without residual
entanglement with the field.

To see this explicitly, we first rederive the effective Hamiltonian~\eq{eq_push_H_simple} in the presence of a
single drive of frequency $\omega$ and of amplitude $\veps(t)$.  Here, we allow the amplitude to be
time-dependent and complex.  Moreover, we will assume that the qubits are detuned from each other
($\wa{1}''\neq\wa{2}''$) such that the flip-flop interaction can be neglected and that both qubits are
dispersively coupled to the resonator.  In a frame rotating at the resonator frequency $\wrr$, we obtain 
\be
\tilde H(t) \approx  \frac{1}{2} \sum_{j=1,2} \left( \wa{j}''(t) + 2f_j(t)\ad + 2f_j^*(t)\aop \right)\sz{j},
\label{eq_H_eff_push}
\ee
where
\be
\begin{split}
& f_j(t)
\\
& = g_j^2 e^{-i \Delta_j t} \int_0^t\int_0^{t'} dt' dt'' e^{i (\Delta_{j} +i\kappa/2)t'} e^{i
(\Delta_r-i\kappa/2) t''} \veps(t'').
\end{split}
\label{eq_f}
\ee
Because of the time dependent drive, the shifted qubit transition frequencies $\wa{j}''(t)$ are now
time-dependent. Note that, following the procedure of appendix~\ref{sec_displace_master}, we have added the
effect of cavity damping $\kappa$ to Eq.~\eq{eq_f}.
%

Since
\be
\begin{split}
&[\tilde H(t_1),\tilde H(t_2)]
\\
& = 2i\left(\im\right[f_1^*(t_1)f_2(t_2)\left] + \im\left[f_2^*(t_1)f_1(t_2)\right] \right)\sz{1}\sz{2}
\\
& \equiv 2iF(t_1,t_2)\sz{1}\sz{2}
\end{split}
\ee
commutes with $\tilde H(t)$ for all times, evolution under the effective Hamiltonian~\eq{eq_H_eff_push} is given
by
\be
U(T,0) = e^{-i\int_0^T dt \tilde H(t)} e^{-\frac{1}{2}\int_0^T\int_0^{t} dtdt' [\tilde H(t),\tilde H(t')]}.
\ee
To avoid unwanted entanglement of the qubits with the resonator at the final gate time $T$, we choose the
time-dependent drive amplitude $\veps(t)$ such that~\cite{garcia-ripoll:2003,garcia-ripoll:2005}
\be
\int_0^T dt f_j(t) =0. \label{eq_int_f_0}
\ee

With this choice of $\veps(t)$, evolution under $\tilde H(t)$ corresponds to the application of a local phase
shift
\be
\phi_j(T) = \int_0^T dt \wa{j}''(t)
\ee
to each qubit and of a conditional phase shift $\exp\left[-i\Phi_{12}(t)\sz{1}\sz{2}/2\right]$ where 
\be
\Phi_{12} (T) = 2 \int_0^T\int_0^{t} dt dt' F(t,t').
\ee
For $\Phi_{12}(T) = \pm\pi/2$, this two-qubit operation is known to be equivalent to the CNOT gate, up to one qubit
gates.

Our goal is therefore to choose the drive $\veps(t)$ such that the qubits accumulate a phase $\Phi_{12} (T) =
\pm\pi/2$ in the smallest time $T$ possible.  This minimization has to take into account several constraints.  In
addition to~\eq{eq_int_f_0}, we take the drive to be off at the start and at the end of the gate:
\be
\veps(0) = \veps(T) = 0. \label{eq_eps_0}
\ee
The assumption that the drive is initially turned off is already built into our expression for $f_j(t)$ in
Eq.~\eq{eq_f}.  To prevent further phase accumulation after the gate time $T$, we also require that
\begin{align}
\alpha(T) &= 0,
\label{eq_alpha_0_constraint}\\
\dot\alpha(T) &=0,
\end{align}
where $\alpha(t)$ is the amplitude of the classical field inside the resonator and is given by
Eq.~\eq{eq_alpha_displacement} [or Eq.~\eq{eq_alpha_dot_with_kappa} in presence of cavity damping].    From
Eq.~\eq{eq_alpha_displacement}, we have that $\dot\alpha(T)=0$ will be automatically satisfied if
Eqs.~\eq{eq_eps_0} and~\eq{eq_alpha_0_constraint} are satisfied.

While satisfying the above constraints, the external field $\veps(t)$ must also be chosen such that the
approximation that led to Eqs.~\eq{eq_H_eff_push} and \eq{eq_f} are valid.  There are two approximations.  The
first one, as in Section~\ref{sec_1q_gates_off_res},  is that we adiabatically eliminated transitions in the
qubit caused by the external drive.  This requires the drive to be either of small amplitude or sufficiently
detuned from the qubits.  The second approximation is the dispersive transformation, which here only has the effect of
renormalizing the qubit transition frequency $\wa{j}''$.   This approximation breaks down on a scale given by
the critical photon number $\nc = \Delta^2/4g^2$~\cite{blais:2004,gambetta:2006}.  We must therefore choose the
drive amplitude and frequency such that the resonator never gets populated by more than $n_\mathrm{crit}$
photons. Beyond this number, additional mixing between the qubits and resonator states is possible and would
likely lead to unwanted qubit-resonator entanglement~\footnote{It is possible that the present approach could
work beyond $n_\mathrm{crit}$ with proper pulse shaping and timing, but investigating this regime requires
extensive numerical calculations that are beyond the scope of this paper.}

Finally, we also require that the measurement-induced dephasing time of the qubits due to the control drive,
the gate-induced dephasing time, to be smaller than the `intrinsic' dephasing time $T_2$.   An expression for the
measurement-induced dephasing rate is given in Eq.~\eq{eq_Gamma_m}.  Here, we will use  
Eq.~(5.16) of Ref.~\cite{gambetta:2006} that is more appropriate for a time-dependent drive.
As was discussed in section~\ref{sec_control_vs_measurement} and in
Ref.~\cite{gambetta:2006}, by working at sufficiently large detuning $\Delta_r$, gate-induced
dephasing can be made negligible while still maintaining a large rate for the gates.

It is useful to take into account these constraints by developing the pulse envelope $\veps(t)$ in Fourier
components
\be
\veps(t) = \sum_{n=-\infty}^\infty c_n e^{+i\w{n} t}.
\ee
Using this expression, we rewrite $f_j(t)$ as
\begin{widetext}
\be
\begin{split}
f_j(t) & = g_j^2 \sum_{n=-\infty}^\infty c_n \left\{ \frac{-e^{i
(\Delta_r+\w{n})t}}{(\Delta_j+\Delta_r+\w{n})(\Delta_r+\w{n}-i\kappa/2)}
+\frac{e^{-\kappa t/2}}{(\Delta_{j}+i\kappa/2)(\Delta_r+\w{n}-i\kappa/2)}
+\frac{-e^{-i \Delta_j t}}{(\Delta_j + \Delta_{r}+\w{n})(\Delta_j+i\kappa/2)} \right\}
\\
&\approx -\chi_j \sum_{n=-\infty}^\infty c_n \frac{e^{i (\Delta_r+\w{n})t}-e^{-\kappa
t/2}}{(\Delta_r+\w{n}-i\kappa/2)}
\end{split}
\label{eq_approx_fj}
\ee
\end{widetext}
where we have assumed the resonator-qubit detuning $\Delta_j$ is large, such that
$\Delta_j\gg\{\Delta_r+\w{n},\kappa/2\}$, and have dropped the small and fast oscillating last term 
in the second expression. This approximate expression will be useful in obtaining analytical estimates 
for the expected gate time T.

Using the approximate expression for $f_j(t)$, we can now recast the above constraints in more simple 
forms.  Using \eq{eq_approx_fj},
the no-effect on the resonator constraint of Eq.~\eq{eq_int_f_0} can be written as
\be
\begin{split}
\int_0^T dt f_j(t) & \approx -\chi_j \sum_{n=-\infty}^\infty c_n \int_0^T dt
\frac{e^{i (\Delta_r+\w{n})t}-e^{-\kappa t/2}}{(\Delta_r+\w{n}-i\kappa/2)}\\
& \equiv \chi_j \sum_{n=-\infty}^\infty c_n A_{n} = 0.
\end{split}
\label{eq_constraint_An}
\ee
Without the approximation for $f_j(t)$, the coefficient $A_n$ depends on the qubit index $j$
which means that there would be one extra constraint to satisfy.

Moreover, to satisfy Eq.~\eq{eq_eps_0}, we take $\omega_n = 2\pi n /T$ which implies that
\be
\veps(0) = \veps(T) = \sum_{n=-\infty}^\infty c_n = 0.
\ee
To satisfy Eq.~\eq{eq_alpha_0_constraint}, we have
\be
\begin{split}
\alpha(T) &= \sum_{n=-\infty}^\infty c_n
\frac{e^{i (\Delta_r+\w{n})T}-e^{-\kappa T/2}}{(\Delta_r+\w{n}-i\kappa/2)}\\
& \equiv  \sum_{n=-\infty}^\infty c_n B_{n}  = 0.
\end{split}
\label{eq_constraint_Bn}
\ee


Casting, in a simple form, the constraints that there is not qubit transition and that the dispersive
approximation holds is more difficult.  For the former case, we require that
\be
\left| f_j(t) e^{-i\Delta_r t}/g_j \right| \sim \left| \frac{g_j \veps(t)}{\Delta_j \Delta_r}\right|
\ee
be smaller than about 1/10 for all time.  In terms of the drive amplitude we thus require
\be
|\veps(t)| \lesssim  \left| \frac{\Delta_j \Delta_r}{10 g_j}\right|. \label{eq_veps_constraint_beta}
\ee
Finally, for the dispersive approximation to hold, we require that the average photon number populating the
resonator be no larger than the critical photon number:
\be
\bar n \approx \frac{\veps^2}{\Delta_r^2} \sim \frac{\Delta_j^2}{4g_j^2}
\ee
or, equivalently,
\be
|\veps(t)| \lesssim  \left| \frac{\Delta_j \Delta_r}{2 g_j}\right|. \label{eq_veps_constraint_ncrit}
\ee
In practice, we therefore only have to deal with Eq.~\eq{eq_veps_constraint_beta} and only with the qubit for
which $|\Delta_j/g_j|$ is smallest.  In the dispersive regime and in the $\kappa=0$ approximation, we thus require $|\veps(t)| \lesssim  \Delta_r$ which means that there is no gate effect at $\Delta_r \rightarrow 0$.

\begin{figure}[tbp]
\centering \includegraphics[width=1\columnwidth]{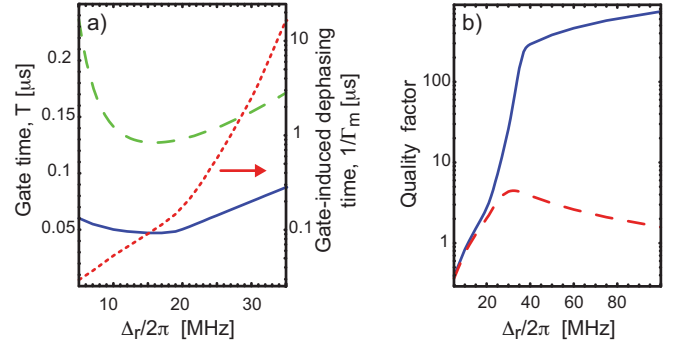}
\caption{ (Color online) a) Gate time (full blue and green dashed) and gate-induced dephasing (red dotted) as a function of detuning $\Delta_r$.  The green dashed line is obtained analytically by decomposing the pulse $\veps(t)$ over four Fourier coefficients and using the constraints derived from the approximate expression in Eq.~\eq{eq_approx_fj}.  The full blue line is obtained numerically by using the constraints derived from the exact expression in Eq.~\eq{eq_approx_fj} and decomposing over 5 Fourier coefficients.  The dotted red line is the gate-induced dephasing time as calculated from Ref.~\cite{gambetta:2006} and presented on a log scale.  The system parameters used are given in the text.  b) Quality factor of the geometric phase gate as a function of detuning.  The red dashed line takes into account gate-induced dephasing and an additional $T_2$ of 500 ns.   For the full blue line, we have taken $1/T_2=0$.  Clearly, gate-induced dephasing is not a limitation in practice.}
\label{fig_geo_gate_time}
\end{figure}

Of all the constraints, the last one is the most difficult to deal with because it must hold at all intermediate
times $t$ and because it involves the absolute value of the complex drive amplitude $\veps(t)$.  
It is nevertheless possible to find an
analytical expression for the gate time $T$ by developing on four Fourier components
$\{c_{-1},c_0,c_1,c_2\}$.   In this situation, all constraints can be solved for analytically and it is possible to find an expression for the gate time.  This expression is however unsightly and we only give here relevant limits.  In the situation where $\Delta_r$ is large, we find that
\be
T 
\sim
\frac{\pi|\Delta_r|}{\chi_1 \chi_2}.
\label{eq_geo_gate_T_bound}
\ee
In the dispersive approximation this is roughly given by $T\sim 100\pi|\Delta_r|/g^2$.  In the small $\Delta_r$ limit we rather find that $T\propto 1/|\Delta_r|$.  As explained above, this behavior is expected from Eq.~\eq{eq_veps_constraint_beta} which is saying that the field amplitude goes to zero at zero detuning in the $\kappa = 0$ limit.  The full expression for $T$ as a function of detuning $\Delta_r$ is plotted in Fig.~\ref{fig_geo_gate_time}a) [dashed green line].  The system parameters used here are are $g_1/2\pi=g_2/2\pi = 100$ MHz, $\Delta_1/2\pi$ = 1 GHz and $\Delta_2/2\pi$ = 1.1 GHz.  The approximate expression for $T$ at large $\Delta_r$  is of similar form to that already obtained for gates based on the dispersive approximation: a rate equal approximately to the square of the coupling $g$ over a detuning.  Here however the detuning is $\Delta_r$ and not the qubit-resonator detuning $\Delta$.  In the dispersive approximation, the former can however be made much smaller than the latter and this gate could in principle be faster than the dispersive based gates discussed in the previous sections. 

Going beyond the assumptions made to obtain the approximate expression for $f_j(t)$ in Eq.~\eq{eq_approx_fj}, we solved numerically the optimization problem.  Without this approximation there are now five constraints to optimize over [the constraint Eq.~\eq{eq_constraint_An} now has to be satisfied independently for both qubits] and thus a minimum of five Fourier coefficients have to be used.  The full blue line in Fig.~\ref{fig_geo_gate_time}a) shows the numerically found gate time as a function of detuning and using the same system parameters as given above.  As can be seen from this figure, going beyond the approximations used to get an analytical estimate and increasing the number of Fourier components improved significantly the gate time.  In this situation, we get an optimal gate time of $T\sim 50$ ns.  Further optimisation can be made by increasing the coupling strength, and could likely be made by  increasing the number of Fourier components.  

Figure~\ref{fig_geo_gate_time}a) also shows the gate-induced dephasing time [dotted red line] associated with the pulse required to implement the geometric phase gate.  This dephasing time is obtained from Eq.~(5.16) of Ref.~\cite{gambetta:2006} and assuming a resonator damping rate of $\kappa/2\pi = 0.1$ MHz.  At detunings larger than about $\Delta_r/2\pi\sim$ 25 MHz, the measurement-induced dephasing time is significantly larger than the `intrinsic' $T_2$ of 500 ns already measured in circuit QED~\cite{wallraff:2005} and can thus be ignored.  However, this induced dephasing time goes down rapidly with detuning such that there is a detuning ($\Delta_r/2\pi \sim 10$ MHz with the chosen parameters) below which it is smaller than the gate time, meaning that the geometric phase gate cannot be used.  

Figure~\ref{fig_geo_gate_time}b) shows the quality factor, as defined by Eq.~\eq{eq_quality_factor} and where the dephasing rate was taken to be the sum of the gate-induced dephasing rate and of the `intrinsic' dephasing rate $1/T_2$.  For the red dashed line, we have taken $T_2$ = 500 ns, while $T_2$ was set to infinity for the full blue line.  For $T_2 = 500$ ns and the present system parameters, the quality factor is maximum for detunings of about 30 MHz.  However, as is clear from the full blue line, gate-induced dephasing time is not a limitation in practice and the quality factor could be significantly better once $T_2$ is improved in this system and by working at larger detunings.  Moreover, the actual magnitude of the quality factor could likely be improved by increasing the number of Fourier components.  
We also point out that the present results have been obtained within the approximations used to derive the effective Hamiltonian of Eq.~\eq{eq_H_eff_push}. Significant improvements could be made by numerically optimizing the full system's master equation. In this case, the results would not be limited by the dispersive approximation used here and we expect a much better quality factor.

Finally, it is important to realize that whenever two qubits are present in the resonator and the system is
being driven, this geometric gate is in action.  This could lead to unwanted qubit-qubit entanglement and, since
Eq.~\eq{eq_int_f_0} might not be satisfied, qubit-resonator entanglement if the drive amplitude and frequency
are not chosen appropriately.

\section{Conclusion}

\begin{table*}
\begin{ruledtabular}
\begin{tabular}{l|c|c|c|c|c}
& Rates & Rates (current) & Q (current)  & Rates (better)  & Q (better)\\\hline 
\S\ref{sec_variable_detuning} Direct coupling& $g$& $\lesssim$ 100& $\lesssim$ 272& $\lesssim$ 200&
$\lesssim$ 619\\
\S\ref{sec_virtual_gate}  Virtual interaction& $g^2/\Delta$& $\sim$ 10& $\sim$ 16& $\sim$ 20&
$\sim$ 31\\
\S\ref{sec_cqed_sidebands} Red/Blue sideband& $g (\Omega_{R_{1}}/2\Delta_{a_{1}})
(\Omega_{R_{2}}/2\Delta_{a_{2}})$& $>$ 1& $>$ 3& $>$ 2&
$>$ 6\\
\S\ref{eq_ac-dither_sideband} Red/Blue sideband (ac-dither)& $g (\Omega_R/2\Delta_a) J_1\!\left(8\ec{}
n_g^{ac}/\ej{}\right)$& $>$ 3& $>$ 8& $>$ 6&
$>$ 19\\
\S\ref{sec_push_gate} Geometric phase& $> g^2/200\Delta_r$& $>$ 3& $>$ 4& $>$ 6& $>$ 9 
\end{tabular}
\end{ruledtabular}
\caption{Rates and quality factors of the various gates.  For the direct coupling gate, the values presented are
upper bounds.  For the sidebands and geometric phase gates the results presented are only {\em lower bounds}.
The system parameters are taken as $g/\Delta = 0.1 = \Omega_R/2\Delta_a$ = 0.1 for the approximations used to
derive the rates to be valid.   For the `current' results, we have taken $g/2\pi$ = 100 MHz and $\kappa/2\pi$ =
0.1 MHz.  For the `better' results, we have taken $g/2\pi$ = 200 MHz and $\kappa/2\pi$ = 0.01 MHz.  The
dephasing time was taken as $T_2 = 500$ ns, as measured in Ref.~~\cite{wallraff:2005} in this system.  For
the geometric phase, we took $\veps = \Delta_r$ to satisfy both Eq.~\eq{eq_veps_constraint_beta} and
the dispersive approximation.  For the ``current" results, we have taken  $\Delta_r/2\pi$ = 32 MHz and for the
``better" results  $\Delta_r/2\pi$ = 70 MHz.}
\label{table_rates}
\end{table*}

We have explored the realization of single and two-qubit gates in circuit QED, using realistic systems
parameters.  We have shown how all single-qubit rotations can be realised with minimal measurement-induced
dephasing of the qubit.  In this context, two approaches to change the qubit transition frequency without dc
bias away from the sweet spot were discussed.  Both rely on off-resonant irradiation of the qubit.
Interestingly, the ac-dither approach could help in reducing dephasing by protecting the qubit from
low-frequency noise.

Five types of two-qubit gates were discussed.  The fastest gate discussed in this paper is based on direct
coupling of qubits with the resonator.    In this paper however, we have focused on gates requiring no dc-bias
away from the sweet spot or additional control lines not present in the current circuit QED architecture.  These
gates can in practice be slower but could be implemented in the current circuit QED design, without additional
design elements.  Moreover, these gates have the important advantage that they do not cause extra dephasing of
the qubits by moving away from the sweet spot.

The rates and quality factors for these gates, as obtained from perturbation theory, are summarized in
Table~\ref{table_rates}.  These are given for $g/2\pi = 100$ MHz and $\kappa/2\pi$ = 0.1 MHz (`current') and
$g/2\pi = 200$ MHz and $\kappa/2\pi$ = 0.01 MHz (`better').  All other parameters are equal and discussed in the
table caption.  These parameters correspond to already realised values and to realistic values that could be
obtained in future realizations of circuit QED, respectively.  The quality factors are calculated in the same
way as in Eq.~\eq{eq_quality_factor}:  the rates divided by $2 T_2^{-1}$ (the full width at half-maximum of the
qubit spectral line).
However, for the direct coupling and the sidebands, the loss of a single photon has a large impact and the
quality factors are therefore given by the rates divided by the mean of the two contributing decay channels:
$(\kappa+2 T_2^{-1})/2$.  For the geometric phase gate, the `current' quality factor, and the corresponding rate,  are taken from the inset of Fig.~\ref{fig_geo_gate_time}.  The `better' results are obtained from a similar calculation.

Moreover, we note that the same values of $T_2$ and $\kappa$ were used for the direct coupling gate requiring dc
excursions away from the sweet spot and the other gates that do not require such excusions.  As discussed
before, it is likely that $T_2$ or $\kappa$ would be reduced due to the dc-bias..  The rates and quality
factors given on the first line of the table can therefore be taken as upper bounds.

It is also important to realise that, appart from the first line of the table, these results have been obtained
by perturbation theory and are thus {\em lower bounds} on the rates and quality factors that can be obtained in
practice.  Better results can be obtained from numerical optimization.  This was  already shown in
Fig.~\ref{fig_sidebands_oscillations} for the red and the blue sideband and also in section~\ref{sec_push_gate}
for the geometric phase.    Therefore, there is good hope that all of the gates discussed in this paper could be realized experimentally, admittedly with different degree of usefulness for quantum information processing.  Moreover, all the quality factors quoted in table \ref{table_rates} are limited by the system's decoherence time $T_2$ and not by measurement-induced dephasing rate from the application of the gate.  As a result, increasing $T_2$ in this system will lead to significantly better quality factors.

The key to improve the gate's quality factor is to improve the coupling coupling $g$, the resonator and qubit
relaxation and dephasing times.  Since the first circuit QED experiment~\cite{wallraff:2004}, $g$ as been
improved by almost a factor of 20 and further improvements can be realized without technical challenges.
Resonators with long photon lifetimes, in the tens of microseconds, have already been
fabricated~\cite{frunzio:2004} and first steps in the design and realization of a new charge-type qubit which is
largely insensitive to charge noise have been taken~\cite{schuster:2007}.   Circuit QED therefore seem like a
promising system with which to study quantum mechanics at the large scales and quantum information processing.

\begin{acknowledgments}
The authors are very grateful to Yale CQUIP visitor Peter Zoller for helpful discussions concerning 
the geometric phase gate and its mapping to circuit QED.
This work was supported in part by the Army Research Office and the National Security Agency through grant
W911NF-05-1-0365, the NSF under grants ITR-0325580, DMR-0342157 and DMR-0603369, and the W. M. Keck Foundation.
AB was partially supported by the Natural Sciences and Engineering Research Council of Canada (NSERC), the Fond
Qu\'eb\'ecois de la Recherche sur la Nature et les Technologies (FQRNT) and the Canadian Institute for Advanced
Research (CIAR).
\end{acknowledgments}

\appendix

\section{Field displacement on the master equation}
\label{sec_displace_master}

In this appendix,  we apply the field displacement procedure of section \ref{sec_review_1_qubit} on the  the
master equation~\eq{eq_master_eq} rather than on the pure state evolution.  Since the displacement is on the
resonator field, we only consider the contribution to damping due to $\kappa$.  The master equation thus reads
\be
\dot\rho = -i[H,\rho] +\frac{\kappa}{2} \left(2 \aop \rho \ad - \ad\aop \rho - \rho \ad\aop
\right),
\ee
where
\be
\begin{split}
H &=\wrr \ad \aop + \frac{\wa{}}{2}\sz{} - g \left(\ad\smm{}+\spp{} \aop\right)
\\
&+\sum_{k} \left(\veps_{k}(t)  \ad e^{-i \wdr{k}t}+ \veps_{k}^*(t)  \aop e^{+i \wdr{k}t}\right).
\end{split}
\ee
For simplicity of notation, in this appendix we only consider a single qubit and drop the $j$ index.  Going to
the displaced frame, the master equation for the displaced density matrix $\tilde\rho = D^\dag(\alpha)\rho
D(\alpha)$ reads
\be
\begin{split}
\dot{\tilde\rho} = & -i [D^\dag(\alpha)HD(\alpha),\tilde\rho] - D^\dag(\alpha)\dot
D(\alpha)\tilde\rho - \tilde\rho \dot D^\dag(\alpha) D(\alpha)
\\
& +\frac{\kappa}{2} D^\dag(\alpha)\left(2 \aop \rho \ad - \ad\aop \rho - \rho \ad\aop \right)D(\alpha)
\\
= & -i [\tilde H , \tilde\rho ] +\frac{\kappa}{2} \left(2 \aop \tilde\rho \ad - \ad\aop \tilde\rho - \tilde\rho
\ad\aop \right)
\end{split}
\ee
where
\be
\begin{split}
\tilde H & = \wrr  \ad \aop + \frac{\wa{}}{2}\sz{} - g \left(\ad\smm{}+\spp{} \aop\right)\\
&- g \left(\alpha^*\smm{}+\alpha\spp{} \right)
\end{split}
\ee
and the parameter $\alpha$ is chosen to satisfy
\be
\dot\alpha = -i\wrr \alpha -i\sum_k \veps_k e^{-i\wdr{k} t} -\frac{\kappa}{2}\alpha.
\label{eq_alpha_dot_with_kappa}
\ee

As an example, we consider the simple case of the bit-flip gate discussed in section~\ref{sec_1_qubit_gate_x}.
In this situation, a single time-independent and real drive $\veps$ is needed.   With $\alpha(0) = 0$ and
dropping the transient term, we recover the Hamiltonian~\eq{eq_Rabi} where the Rabi rate now reads
\be
\Omega_R = 2 \frac{g\veps}{\Delta_r - i \kappa/2}.
\ee
As it should, this rate does not diverge at $\Delta_r=0$.  Since we always work in the regime where $\Delta_r$
is large, we neglect this $\kappa$ correction in most sections of this paper.

\section{Derivation of the effective Hamiltonian}
\label{sec_appendix_eff_H}

In this appendix we derive an effective Hamiltonian for two qubits dispersively coupled to a single resonator
and taking into account the presence of three independent drives.  Two of these drives ($k=1,2$) are used to
coherently control the qubits while the third drive ($k=3$) plays the role of measurement beam.  Our starting
point is Hamiltonian~\eq{eq_H_2qubit_full} taking into account the gate charge dependence.  While working away
from charge degeneracy is not useful in practice, this will make selection rules appear clearly.

The derivation follows the same steps as those presented in section~\ref{sec_review_1_qubit}.  For simplicity,
we take the drive amplitudes $\veps_k$ to be time-independent.  This assumption is relaxed in the context of the
geometric phase gate in section~\ref{sec_push_gate}.  As in section~\ref{sec_review_1_qubit}, we start by
displacing the resonator field $a$ by using the displacement operator $D(\alpha)$.  Since $\veps_3$ plays the
role of measurement, it's frequency will be close to the resonator frequency $\omega_r$.  It is therefore more
convenient to displace the field only with respect to the first two drives.  In the lab frame, the result of
this transformation is
\begin{equation}
\begin{split}
H^{(1)} & = \omega_r \ad \aop + \frac{\wa{1}}{2}\sz{1} + \frac{\wa{2}}{2}\sz{2}
+ \veps_3\left(\ad e^{-i\wdr{3} t} + \mathrm{h.c.} \right) 
\\
& - \sum_{j=1,2} g_j \left( \mu_j - \ccos_j \sz{j} + \ssin_j\sx{j}  \right)
\left(\ad+\aop\right)\\
& +\sum_{j,k=1,2} \Omega_{R_{jk}} \left( \ccos_j \sz{j} - \ssin_j\sx{j}  \right) \cos(\wdr{k} t),
\end{split}
\label{eq_ap_H_1}
\end{equation}
where $\Omega_{R_{jk}} = 2 g_j\veps_k/(\omega_r-\wdr{k})$ is the Rabi frequency of qubit $j$ with respect to
drive $k$.

In the next step, we make the rotating wave approximation on the drive terms acting on the qubits [last term of
Eq.~\eq{eq_ap_H_1}]:
\begin{equation}
\begin{split}
&\sum_{j,k=1,2} \Omega_{R_{kj}} \left( \ccos \sz{j} - \ssin\sx{j}  \right) \cos(\wdr{k} t)
\\
&\approx \sum_{j,k=1,2} \ssin_j\frac{\Omega_{R_{kj}}}{2} \left( \spp{j}e^{-i\wdr{k} t} + \smm{j}e^{+i\wdr{k} t}
\right).
\end{split}
\end{equation}
Not doing this approximation would only lead to small Bloch-Siegert shifts in the qubit transition
frequency~\cite{bloch:1940}.

In this appendix, we are not interested in direct transitions of the qubits (i.e single-qubit Rabi flopping),
these transitions are therefore eliminated using the transformation of Eq.~\eq{eq_U_off_qubit} on each qubit:
\be
U_j=\exp\left[\beta_j^*\spp{j}-\beta_j\smm{j}\right]. \label{eq_U_qubit_drive_elimination}
\ee
Since $[U_1,U_2]=0$, these two transformations can be applied sequentially:
\begin{equation}
H^{(2)}
 =  (U_1U_2)H^{(1)} (U_1U_2)^\dag -  i \sum_{j=1,2} U_j\dot U_j^{-1}.
\label{eq_H_2}
\end{equation}
We expand the first term to second order in the small parameter $\beta_j$ using the Hausdorff formula
Eq.~\eq{eq_Hausdorff}.  For the second term of Eq.~\eq{eq_H_2}, we obtain to second order:
\begin{equation}
\begin{split}
U_j\dot U_j^{-1} \approx
 \frac{1}{2}(\beta_j^*\dot\beta_j-\beta_j\dot\beta_j^*)\sz{j}
+ (\dot\beta_j\smm{j}-\dot\beta_j^*\spp{j}).
\end{split}
\end{equation}
Choosing
\be
\beta_j(t) =  \frac{1}{2}\sum_{k=1,2}\frac{\Omega_{R_{jk}}}{\Delta_{a_{jk}}}e^{+i\wdr{k}t}
\ee
with $\Delta_{a_{jk}}=\wa{j}-\wdr{k}$ and neglecting fast oscillating terms, we finally obtain
\begin{equation}
\begin{split}
H^{(2)}  & \approx \omega_r \ad\aop +\sum_{j=1,2} \frac{\wa{j}'}{2}\sz{j}
+ \veps_3\left(\ad e^{-i\omega_3 t} + \mathrm{h.c.} \right)\\
&
-\sum_{j=1,2} g_j \left(\mu_j+\ssin_j\sx{j}\right)\left(\ad+\aop\right)\\
&
-\sum_{j=1,2} g_jA_j \left(\beta_j^*\spp{j}+\beta_j\smm{j} \right)\left(\ad+\aop\right)\\
& -\sum_{j=1,2} g_jB_j \sz{j}\left(\ad+\aop\right),
\end{split}
\end{equation}
where the shifted qubit transition frequency is
\be
\begin{split}
\wa{j}' & = \wa{j}
+ \sum_{k=1,2} \frac{\Omega_{R_{jk}}^2}{2\Delta_{a_{jk}}} \\
&+ \frac{\Omega_{R_{j1}}\Omega_{R_{j2}}(\Delta_{a_{j1}}+\Delta_{a_{j2}})}{2\Delta_{a_{j1}}\Delta_{a_{j2}}}
\cos(\wdr{1}-\wdr{2})t
\end{split}
\label{eq_wa_shift_2_drives}
\ee
and
\be
\begin{split}
A_j &= 2\ccos_j-\ssin_j(\beta_j^*+\beta_j)\\
B_j &= 2\ccos_j\left(|\beta_j|^2-\frac{1}{2}\right)+\ssin_j(\beta_j^*+\beta_j).
\end{split}
\label{eq_B}
\ee
In the presence of a single off-resonant drive, we recover Eq.~\eq{eq_phase_gate_1q}.

Finally, the qubits are assumed to be strongly detuned from the resonator.  We therefore adiabatically eliminate
the direct Jaynes-Cummings qubit-resonator interaction.  This is done using the dispersive transformation of
Eq.~\eq{eq_U_dispersive_2q}.  Since the rotating wave approximation was not performed on the qubit-resonator
interaction, this choice of transformation will not cancel completely the interaction.  Complete cancelation
could be obtained by choosing a slightly different transformation.  However, for the transitions of interest
here, the remaining terms will be oscillating rapidly and will simply be dropped.  Taking into account these
terms would, again, only add a small frequency shift~\cite{bloch:1940} that can safely be ignored here.

Applying the dispersive transformation Eq.~\eq{eq_U_dispersive_2q}, expanding to second order in the small
parameter $g_j/\Delta'_j$ and neglecting fast oscillating terms, we obtain the main result of this appendix
\begin{widetext}
\be
\begin{split}
H^{(3)}  & \approx \omega_r \ad\aop + \sum_{j=1,2}\frac{\wa{j}''}{2}\sz{j} + \veps_3\left(\ad e^{-i\wdr{3} t}
+\aop e^{+i\wdr{3} t} \right)
-\sum_{j=1,2}\frac{\veps_3}{2}\left(\frac{g_j\ssin_j}{\Delta'_j}\right)^2 \left(\ad e^{-i\wdr{3} t}  +\aop
e^{+i\wdr{3} t} \right)\sz{j}
\\
& -\sum_{j=1,2}g_jB_j\sz{j}(\ad+\aop) +\ssin_1\ssin_2\frac{g_1g_2(\Delta'_1+\Delta'_2)}{2\Delta'_1\Delta'_2}
\left(\spp{1}\smm{2}+\smm{1}\spp{2}\right)
\\
& -\sum_{j,k=1,2}\ccos_jg_j\frac{\Omega_{R_{jk}}}{\Delta_{a_{jk}}}
\left( \spp{j}e^{-i\wdr{k} t} + \smm{j}e^{+i\wdr{k} t}  \right)\left(\ad+\aop\right)\\
& +\sum_{j,k,k'=1,2}\ssin_jg_j \frac{\Omega_{R_{jk}}}{2\Delta_{a_{jk}}}
\frac{\Omega_{R_{jk'}}}{2\Delta_{a_{j{k'}}}}
\left\{\left(e^{+i(\wdr{k}-\wdr{k'})t}+e^{-i(\wdr{k}+\wdr{k'})t}\right)\spp{j} +
\left(e^{-i(\wdr{k}-\wdr{k'})t}+e^{+i(\wdr{k}+\wdr{k'})t}\right)\smm{j} \right\}(\ad+\aop),
\end{split}
\label{eq_H_eff_all}
\ee
\end{widetext}
where the shifted qubit transition frequency is
\be
\wa{j}''=\wa{j}'+2\ssin_j^2\frac{g_j^2}{\Delta'_j}
\left(\ad\aop+\frac{1}{2}\right) \label{eq_twice_shifted_wa}
\ee
and we have defined
\be
\Delta'_j=\wa{j}'-\omega_r.
\ee
This effective Hamiltonian contains all the physics needed to realize each of
the different gates that are studied in this paper. More particularly, the first term in the second line
of~\eq{eq_H_eff_all} can be used to generate a geometric two-qubit phase
gate~\cite{garcia-ripoll:2003,garcia-ripoll:2005} and is studied in more details in section~\ref{sec_push_gate}.
The second term of the second line of equation~\eq{eq_H_eff_all} is the flip-flop interaction due to virtual
interaction with the resonator and is discussed in section~\ref{sec_virtual_gate}. We note that higher order
flip-flop terms induced by the external drives have been dropped.  Finally, as discussed in
section~\ref{sec_cqed_sidebands}, the last two lines describe one and two photon blue and red sideband
transitions.

Higher order terms in the perturbative expansions used to obtain Eq.~\eq{eq_H_eff_all} will lead to
additional non-linear terms.  These corrections will be negligible as long as the parameters $\beta_j$
and $g_j/\Delta_j$ are chosen to be small.  This can be safely done in the context of the quantum gates 
studied here.  The situation where these terms can no longer be neglected will be discussed elsewhere.

\section{Red and Blue Sidebands using FLICFORQ}
\label{sec_appendix_flicforq_1q}

Here we consider a single qubit coupled to the resonator and in the presence of a single drive of frequency
$\wa{}$ and amplitude $\veps$.  For simplicity, we work at charge degeneracy and in the rotating
wave-approximation:
\begin{equation}
\begin{split}
H =\: & \omega_r \ad \aop + \frac{\omega_a}{2}\sz{} - g(\ad \smm{} + \spp{} \aop)\\
&+ \veps (\ad e^{-i \omega_a t} + \aop e^{i \omega_a t}).
\end{split}
\label{eq_full_cavity_H}
\end{equation}
We again assume that the qubit is far detuned from the resonator frequency.  However, we choose not to
adiabatically eliminate the qubit-resonator interaction and look at a non-perturbative result for large drive
amplitude.  To find the effective qubit-resonator Hamiltonian in this case, we move to a frame where a
non-vanishing interaction remains, but only when a resonance condition to be determined is satisfied.

We first move to a frame rotating at the qubit transition frequency $\wa{}$ for both the qubit and the
resonator:
\begin{equation}
H^{(1)} = -\Delta \ad \aop - g(\ad \smm{} + \spp{} \aop) + \veps (\ad+ \aop),
\end{equation}
with $\Delta=\wa{}-\wrr$.  We then displace the resonator field using the displacement operator $D(\alpha)$ to
obtain
\begin{equation}
H^{(2)} = -\Delta \ad \aop - g(\ad \smm{} + \spp{} \aop) - \frac{\Omega_R}{2} \sx{}. \label{eq_H_displaced}
\end{equation}
For convenience, we now change $\sx{}$ to $-\sz{}$ using a rotation along the $y$ axis of angle $-\pi/2$.  After
that rotation, the system looks like a qubit of frequency $\Omega_R$ coupled to a resonator of  frequency
$\Delta$:
\begin{equation}
H^{(3)} = -\Delta \ad \aop + \frac{\Omega_R}{2} \sz{} + \frac{g}{2}(\ad + \aop)\sz{} + i\frac{g}{2} (\ad -
\aop)\sy{}.
\end{equation}
Going to a frame rotating at both the new effective qubit and cavity frequencies, we obtain a single
non-oscillating term when $\Delta = -\Omega_R$:
\begin{equation}
H^{(4)}_\mathrm{r} \approx -\frac{g}{2}(\ad \smm{} +  \spp{} \aop).
\end{equation}
and for $\Delta = \Omega_R$ we obtain:
\begin{equation}
H^{(4)}_\mathrm{b} \approx \frac{g}{2}(\ad \spp{} +  \smm{} \aop).
\end{equation}
These correspond, in the rotating frame, to red and blue sideband transitions at a rate $g/2$.  The factor of
$1/2$ is due to the fact that only one of the Rabi sidebands for the ground and excited states of the qubit are
in resonance at $\Delta = \pm\Omega_R$.
This type of qubit-cavity resonance was also discussed in Refs.~\cite{Jonatha:2000} and
\cite{solano:2003} and for flux qubits in a dressed basis in Ref~\cite{liu:2006}.

\section{FLICFORQ with two qubits}
\label{sec_appendix_flicforq_2q}

To obtain the effective Hamiltonian for the two-qubit FLICFORQ, we follow the results of
appendix~\ref{sec_appendix_eff_H} and of Ref.~\cite{rigetti:2005}.  The starting point is the
Hamiltonian~\eq{eq_ap_H_1} taken in the rotating wave approximation and at charge degeneracy for both qubits
($\ssin_j=1, \ccos_j=0$).  Moreover, we omit the measurement drive and consider only two coherent drives
($k=1,2$).  To get maximal splitting of the Rabi sidebands, the frequency of these drive are chosen such that
$\wdr{j}$ is close to $\wa{j}$.

To derive the effective Hamiltonian, we follow the same steps as in appendix~\ref{sec_appendix_eff_H}.  The main
difference is that in the second step, because of our choice of drive frequencies, we do not adiabatically
eliminate both drives from both qubits.  We adiabatically eliminate only the effect of $\wdr{1}$ on the second
qubit and the effect of $\wdr{2}$ on the first qubit.  The resulting effective Hamitonian is
\be
\begin{split}
H & \approx \wrr{}\ad\aop +\sum_{j=1,2}\frac{\wa{j}''}{2}\sz{j}\\
&+\sum_{j=1,2}\frac{\Omega_{R_j}}{2} \left(\spp{j}e^{-i\wdr{j} t}+\smm{j}e^{+i\wdr{j} t}\right)\\
&+\frac{g_1g_2(\Delta'_1+\Delta'_2)}{2\Delta'_1\Delta'_2} \left(\spp{1}\smm{2}+\smm{1}\spp{2}\right),
\end{split}
\ee
where $\Delta'_j=\wa{j}'-\omega_r$ and
\be
\wa{1(2)}'  = \wa{1(2)}  + 2 \frac{\Omega_{R_{12(21)}}^2}{\Delta_{a_{12(21)}}}.
\ee

Assuming for simplicity that $\wdr {j}=\wa{j}''$, we first go to a frame rotating at $\wdr{j}$ for qubit $j$.
Following ref.~\cite{rigetti:2005}, we then go to a frame rotating along the $x$ axis and at a frequency
$\Omega_{R_j}/2$ for qubit $j$.  This yields
\be
\begin{split}
H &\approx \wrr{}\ad\aop +
\frac{g_1g_2(\Delta'_1+\Delta'_2)}{8\Delta'_1\Delta'_2}\\
& \times \left\{ \left[\sx{1}-i\sin(\Omega_{R_1}t)\sz{1}+i\cos(\Omega_{R_1}t)\sy{1}\right]
\right.\\
& \times \left[\sx{2}-i\sin(\Omega_{R_2}t)\sz{2}-i\cos(\Omega_{R_2}t)\sy{2}\right] e^{+i(\wdr{1}-\wdr{2})t}
\\
& \left. + \mathrm{h.c.} \right\}.
\end{split}
\ee
We now look for terms that do not average to zero.  These will correspond to resonance in the coupled driven
system.  There are several choices of resonances here and, as an example, we choose $\wdr{2}-\wdr{1} =
\Omega_{R_1}+\Omega_{R_2}$.  Using this condition, all terms but a single term are oscillating rapidly and
average out to zero.  The resulting effective Hamiltonian in the quadruply-rotating frame is
\be
H_\mathrm{FF} \approx \wrr \ad \aop + \frac{g_1g_2(\Delta'_1+\Delta'_2)}{16\Delta'_1\Delta'_2}
\left(\sy{1}\sy{2}+\sz{1}\sz{2}\right).
\ee
Other choices of resonances lead to different symmetry for the effective Hamiltonian.  These other possible
coupling Hamiltonians will be discussed elsewhere.

\section{Symmetry and selection rules}
\label{sec_symmetry}

At the charge degeneracy point, the Hamiltonian~\eq{eq_H_2qubit_full} is even in the number of creation and
annihilation operators.  As a result, $C\equiv \ad a+ \sz{}/2$ commutes with the Hamiltonian and the total
excitation number is a conserved quantity.  On the other hand, away from charge degeneracy, in addition to the
regular Jaynes-Cummings term, we get extra terms of the form $g(\mu+\ccos\sz{})(\ad+a)$ which are of odd parity
in the number of creation and annihilation operators.  Clearly,  $C$ is not a conserved quantity in this case.

These symmetry considerations will impose selection rules on the transitions that are possible.  To see this
more explicitly, we introduce the parity operator $P= e^{-i\pi \ad \aop}\sz{}$ which is the natural unitary
extension of $C$~\cite{lo:1998,talab:2005}.  By writing the parity operator in the form $P=
\sum_{n=0}^\infty(-1)^n \ket{n}\bra{n}\sz{}$, it is simple to verify that $P$ anti-commutes with the drive
Hamiltonian~\eq{eq_H_drive} in a frame rotating at the drive frequency.  Labeling states as being of even (odd)
parity if they are eigenstates of $P$ with +1 (-1) eigenvalue, it is then clear that the drive $(\ad+\aop)$ can
only cause transitions between states of different parities.

To see which transitions are allowed at the sweet spot, we first give the parity of the eigenstates of the
Jaynes-Cummings Hamiltonian:
\be
P\ket{g0} = -\ket{g0}, \qquad P\ket{\overline{\pm,n}} = (-1)^n \ket{\overline{\pm,n}},
\ee
where~\cite{blais:2004}
\begin{eqnarray}
\ket{\overline{+,n}}
&=& \cos\theta_n \ket{e,n} + \sin\theta_n \ket{g,n+1}\\
\ket{\overline{-,n}} &=& -\sin\theta_n \ket{e,n} + \cos\theta_n \ket{g,n+1}
\end{eqnarray}
and
\be
\theta_n = \frac{1}{2}\tan^{-1}\left(\frac{2g\sqrt{n+1}}{\Delta}\right).
\ee

The red sideband transition illustrated in Fig.~\ref{fig_red_blue}a) corresponds to a transition between
$\ket{\overline{+,0}}$ and $ \ket{\overline{-,0}}$, it is therefore forbidden to first order.  This is the
result already obtained in Eq.~\eq{eq_Heff_1_photon_red}.  In the same way, the blue sideband corresponds to a
transition between $\ket{g0}$ and $\ket{\overline{-,1}}$.  It is also forbidden at charge degeneracy.
Single-photon Rabi flopping discussed in section~\ref{sec_1_qubit_gate_x} connects $\ket{g0}$ and
$\ket{\overline{-,0}}$, and it is obviously allowed at charge degeneracy.  In the case of two photon
transitions, the drive Hamiltonian is effectively acting twice and the previous selection rules are therefore
simply reversed.  Finally, away from the charge degeneracy point, the Hamiltonian has no definite parity and
these considerations do not apply.  All sideband transitions are therefore allowed.  Similar selection rules
for flux qubits coupled to a LC oscillator were studied in Ref.~\cite{liu:2005a}.  In Ref.~\cite{liu:2005}, these
selection rules were investigated for a flux qubit irradiated with classical microwave signal.



\end{document}